\documentclass[twocolumn]{aastex631}
\submitjournal{AJ}

\usepackage[utf8]{inputenc} 
\usepackage{amsmath}
\usepackage{multirow}

\newcommand\tess{TESS}
\newcommand\gaia{\textit{Gaia}}

\newcommand\ms{$\textrm{m~s}^{-1}$}

\newcommand\gcmcubed{$\textrm{g~cm}^{-3}$}

\newcommand{\unit}[1]{\ensuremath{\, \mathrm{#1}}} 
\newcommand\earthmass{$M_{\oplus}$}
\newcommand\earthradius{$R_{\oplus}$}

\hypersetup{linkcolor=violet,citecolor=purple}

\usepackage[inline]{enumitem}

\begin{document}

\title{\textit{Searching for GEMS:} Two Super-Jupiters around M-dwarfs -- Signatures of Instability or Accretion?}


\author[0009-0000-1825-4306]{Andrew Hotnisky}
\affil{Department of Astronomy \& Astrophysics, 525 Davey Laboratory, The Pennsylvania State University, University Park, PA 16802, USA}
\affil{Center for Exoplanets and Habitable Worlds, 525 Davey Laboratory, The Pennsylvania State University, University Park, PA 16802, USA}

\author[0000-0001-8401-4300]{Shubham Kanodia}
\affil{Earth and Planets Laboratory, Carnegie Science, 5241 Broad Branch Road, NW, Washington, DC 20015, USA}

\author[0000-0002-2990-7613]{Jessica Libby-Roberts}
\affil{Department of Astronomy \& Astrophysics, 525 Davey Laboratory, The Pennsylvania State University, University Park, PA 16802, USA}
\affil{Center for Exoplanets and Habitable Worlds, 525 Davey Laboratory, The Pennsylvania State University, University Park, PA 16802, USA}

\author[0000-0001-9596-7983]{Suvrath Mahadevan}
\affil{Department of Astronomy \& Astrophysics, 525 Davey Laboratory, The Pennsylvania State University, University Park, PA 16802, USA}
\affil{Astrobiology Research Center, 525 Davey Laboratory, The Pennsylvania State University, University Park, PA 16802, USA}
\affil{Center for Exoplanets and Habitable Worlds, 525 Davey Laboratory, The Pennsylvania State University, University Park, PA 16802, USA}

\author[0000-0003-4835-0619]{Caleb I. Ca\~nas}
\altaffiliation{NASA Postdoctoral Fellow}
\affiliation{NASA Goddard Space Flight Center, 8800 Greenbelt Road, Greenbelt, MD 20771, USA }

\author[0000-0002-5463-9980]{Arvind F.\ Gupta}
\affil{U.S. National Science Foundation National Optical-Infrared Astronomy Research Laboratory, 950 N. Cherry Ave., Tucson, AZ 85719, USA}

\author[0000-0002-7127-7643]{Te Han}
\affil{Department of Physics \& Astronomy, The University of California, Irvine, Irvine, CA 92697, USA}

\author{Henry A. Kobulnicky}
\affiliation{Department of Physics \& Astronomy, University of Wyoming, Laramie, WY 82070, USA}

\author[0000-0002-2401-8411]{Alexander Larsen}
\affiliation{Department of Physics \& Astronomy, University of Wyoming, Laramie, WY 82070, USA}

\author[0000-0003-0149-9678]{Paul Robertson}
\affil{Department of Physics \& Astronomy, University of California Irvine, Irvine, CA 92697, USA}

\author[0009-0009-4977-1010]{Michael Rodruck}
\affil{Department of Physics, Engineering, and Astrophysics, Randolph-Macon College, Ashland, VA 23005, USA}

\author[0000-0001-7409-5688]{Gudmundur Stefansson}
\affil{Anton Pannekoek Institute for Astronomy, University of Amsterdam, Science Park 904, 1098 XH Amsterdam, The Netherlands}


\author[0000-0001-9662-3496]{William D. Cochran}
\affil{McDonald Observatory and Department of Astronomy, The University of Texas at Austin}
\affil{Center for Planetary Systems Habitability, The University of Texas at Austin}

\author[0000-0003-1439-2781]{Megan Delamer}
\affil{Department of Astronomy \& Astrophysics, 525 Davey Laboratory, The Pennsylvania State University, University Park, PA 16802, USA}
\affil{Center for Exoplanets and Habitable Worlds, 525 Davey Laboratory, The Pennsylvania State University, University Park, PA 16802, USA}

\author[0000-0002-2144-0764]{Scott A. Diddams}
\affil{Electrical, Computer \& Energy Engineering, University of Colorado, 425 UCB, Boulder, CO 80309, USA}
\affil{Department of Physics, University of Colorado, 2000 Colorado Avenue, Boulder, CO 80309, USA}

\author[0000-0002-3853-7327]{Rachel B. Fernandes}
\altaffiliation{President's Postdoctoral Fellow}
\affil{Department of Astronomy \& Astrophysics, 525 Davey Laboratory, The Pennsylvania State University, University Park, PA 16802, USA}
\affil{Center for Exoplanets and Habitable Worlds, 525 Davey Laboratory, The Pennsylvania State University, University Park, PA 16802, USA}

\author[0000-0003-1312-9391]{Samuel Halverson}
\affil{Jet Propulsion Laboratory, 4800 Oak Grove Drive, Pasadena, CA 91109, USA}

\affil{Center for Exoplanets and Habitable Worlds, 525 Davey Laboratory, The Pennsylvania State University, University Park, PA 16802, USA}

\author[0000-0003-1263-8637]{Leslie Hebb}
\affiliation{Department of Physics, Hobart and William Smith Colleges, 300 Pulteney Street, Geneva, NY, 14456, USA}


\author[0000-0002-9082-6337]{Andrea S.J.\ Lin}
\affil{Department of Astronomy \& Astrophysics, 525 Davey Laboratory, The Pennsylvania State University, University Park, PA 16802, USA}
\affil{Center for Exoplanets and Habitable Worlds, 525 Davey Laboratory, The Pennsylvania State University, University Park, PA 16802, USA}

\author[0000-0002-0048-2586]{Andrew Monson}
\affil{Steward Observatory, The University of Arizona, 933 N.\ Cherry Avenue, Tucson, AZ 85721, USA}

\author[0000-0001-8720-5612]{Joe P.\ Ninan}
\affil{Department of Astronomy and Astrophysics, Tata Institute of Fundamental Research, Homi Bhabha Road, Colaba, Mumbai 400005, India}




\author[0000-0001-8127-5775]{Arpita Roy}
\affiliation{Astrophysics \& Space Institute, Schmidt Sciences, New York, NY 10011, USA}

\author[0000-0002-4046-987X]{Christian Schwab}
\affil{School of Mathematical and Physical Sciences, Macquarie University, Balaclava Road, North Ryde, NSW 2109, Australia}

\begin{abstract}
We present the discovery of TOI-6303b and TOI-6330b, two massive transiting super-Jupiters orbiting a M0 and a M2 star respectively, as part of the \textit{Searching for GEMS} survey. These were detected by TESS and then confirmed via ground-based photometry and radial velocity observations with the Habitable-zone Planet Finder (HPF). TOI-6303b has a mass of 7.84 $\pm$ 0.31 M$_J$, a radius of 1.03 $\pm$ 0.06 R$_J$, and an orbital period of 9.485 days. TOI-6330b has a mass of 10.00 $\pm$ 0.31 M$_J$, a radius of 0.97 $\pm$ 0.03 R$_J$, and an orbital period of 6.850 days. We put these planets in context of super-Jupiters around M-dwarfs discovered from radial-velocity surveys, as well as recent discoveries from astrometry. These planets have masses that can be attributed to two dominant planet formation mechanisms --- gravitational instability and core-accretion.  Their masses necessitate massive protoplanetary disks that should either be gravitationally unstable, i.e. forming through gravitational instability, or be amongst some of the most massive protoplanetary disks to form objects through core-accretion. We also discuss the eccentricity distribution of these objects, as a potential indicator of their formation and evolutionary mechanisms.


\end{abstract}



\section{Introduction}\label{sec:intro}

M-dwarfs are the most abundant stars in the Milky Way \citep{henry_solar_2006, reyle_10_2021}. Searches for habitable and Earth-like worlds largely focus on these systems due to the close-in nature of their habitable zones \citep[HZs;][]{kasting_habitable_1993, scalo_m_2007, kopparapu_habitable_2013, childs_life_2022}, and relatively higher planetary detection amplitudes through deeper transit dips, larger RV semi-amplitudes, and more advantageous star-planet contrast ratios when compared to FGK stars. Previous efforts have provided occurrence rates of terrestrial planets around these types of stars \citep{dressing_occurrence_2015, hardegree-ullman_kepler_2019, hsu_occurrence_2020, ment_occurrence_2023}, even though the occurrence rate of habitable-zone planets around M-dwarf stars is still debated \citep{bergsten_no_2023}.


HZ/terrestrial planets can be affected by their giant planet companions \citep{childs_giant_2019, schlecker_new_2021, bitsch_giants_2023}. Around M-dwarfs, giant and terrestrial planets have yet to be seen in the same system. This is derived from a lack of observed giant exoplanets around M-dwarf stars (GEMS). An increase in observed GEMS with precise mass measurements and orbital characteristics would enhance our ability to determine the frequency of giant planets around M-dwarfs and the effects terrestrial and giant planets have on one another. Additionally, we would increase our understanding of giant planet formation around M-dwarfs. 

Core-accretion is the commonly accepted formation pathway for close-in giant planets ($\gtrsim$ 8 \earthradius) \citep{fischer_planet-metallicity_2005, narang_properties_2018}. This formation is a `bottom-up' approach to planetary formation where an $\sim$ 10 M$_\oplus$ core is accreted followed by an exponential gas accretion that forms the planet's gaseous envelope. The alternate mode of formation is the gravitational instability scenario \citep{boss_giant_1997, boss_rapid_2006}, which occurs in the protostellar phase of massive disks where the disk fragments into gravitationally bound clumps. It is expected that core-accretion should produce fewer GEMS because of lower disk masses and longer orbital timescales, causing it to be more difficult to reach the exponential runaway gas accretion \citep{laughlin_core_2004, ida_toward_2005}. NASA's  Transiting Exoplanet Survey Satellite \citep[TESS;][]{ricker_transiting_2014} has enabled the discovery of many  Jupiter-sized exoplanets \citep[e.g.][]{canas_warm_2020,  kanodia_toi-5205_2023, hobson_toi-3235_2023, delamer_toi-4201_2024}, and brown dwarfs \citep[e.g.][]{artigau_toi-1278_2021, canas_eccentric_2022} orbiting M-dwarf stars, which have already started to raise interesting questions about the formation and evolution of these planets \citep{durisen_gravitational_2007, chabrier_giant_2014}. However, the current sample size ($\sim$ 30) is insufficient to enable a comprehensive understanding of these systems. We have therefore started the volume limited ($<$ 200 pc) \textit{Searching for GEMS} survey \citep{kanodia_searching_2024}, to increase the sample size of these planets to $\sim$ 40 transiting GEMS. This will enable robust statistical comparisons of these planets with their FGK counterparts, thereby helping shed light on their formation mechanisms.


Studies of giant planet dependence on host star metallicity have shown that the region between $\sim$ 4 and 10 M$_{J}$ is sparsely populated by giant planets for FGK host stars \citep{santos_observational_2017, schlaufman_evidence_2018, narang_properties_2018}. It is believed that this marks the transition between the two giant planet formation mechanisms --- core accretion and gravitational instability. This is because the objects below 4 M$_{J}$ are believed to have formed through core-accretion, whereas those above 10 M$_{J}$ are formed through gravitational instability \citep{santos_observational_2017, schlaufman_evidence_2018, narang_properties_2018}. 

As part of the \textit{Searching for GEMS} survey, we present the discovery of two super-Jupiters around early M-dwarfs --- TOI-6303b at 7.84 $\pm$ 0.31 M$_J$ and TOI-6330b at 10.00 $\pm$ 0.32 M$_J$. These objects populate the upper limit of the transition zone between core accretion and gravitational instability. In Section~\ref{sec:obs} we present our observations of both systems. We detail our photometric observations from TESS, RBO, APO/ARCTIC, and Keeble observatory, precision radial velocities from HPF, and high-contrast speckle imaging from ShaneAO and NESSI. In Section~\ref{sec:Analysis} we detail the analysis performed to obtain the stellar parameters and the planetary parameters. In Section~\ref{sec:Discussion} we discuss the possible formation paths of these objects and place them in context of the giant exoplanets and brown dwarfs. Lastly, we summarize our discovery in Section~\ref{sec:Conclusion}.

\section{Observations} \label{sec:obs}
\subsection{TESS}
TOI-6303 (TIC-186810676) and TOI-6330 (TIC-308120029) were observed in Sector 18 at 1800-second cadence from 2019 November 3 to 2019 November 27 and in Sector 58 at 200-second cadence from 2022 October 29 to 2022 November 26. Both planet candidates were flagged using the Quick Look Pipeline \citep[QLP;][]{huang_photometry_2020} through the Faint-Star Search \citep{kunimoto_predicting_2022}.

\begin{figure}
    \centering
    \includegraphics[width=\columnwidth]{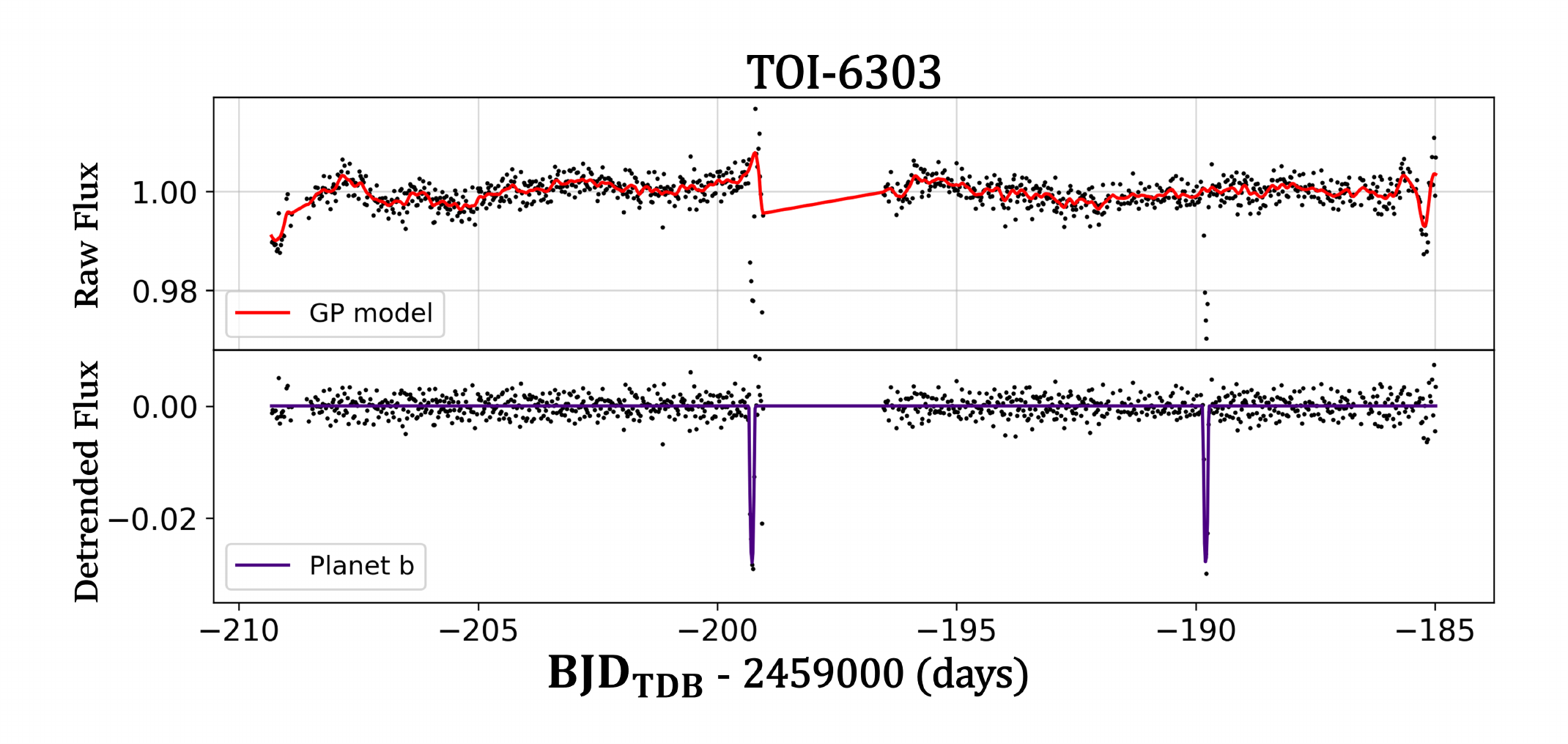}
    \includegraphics[width=\columnwidth]{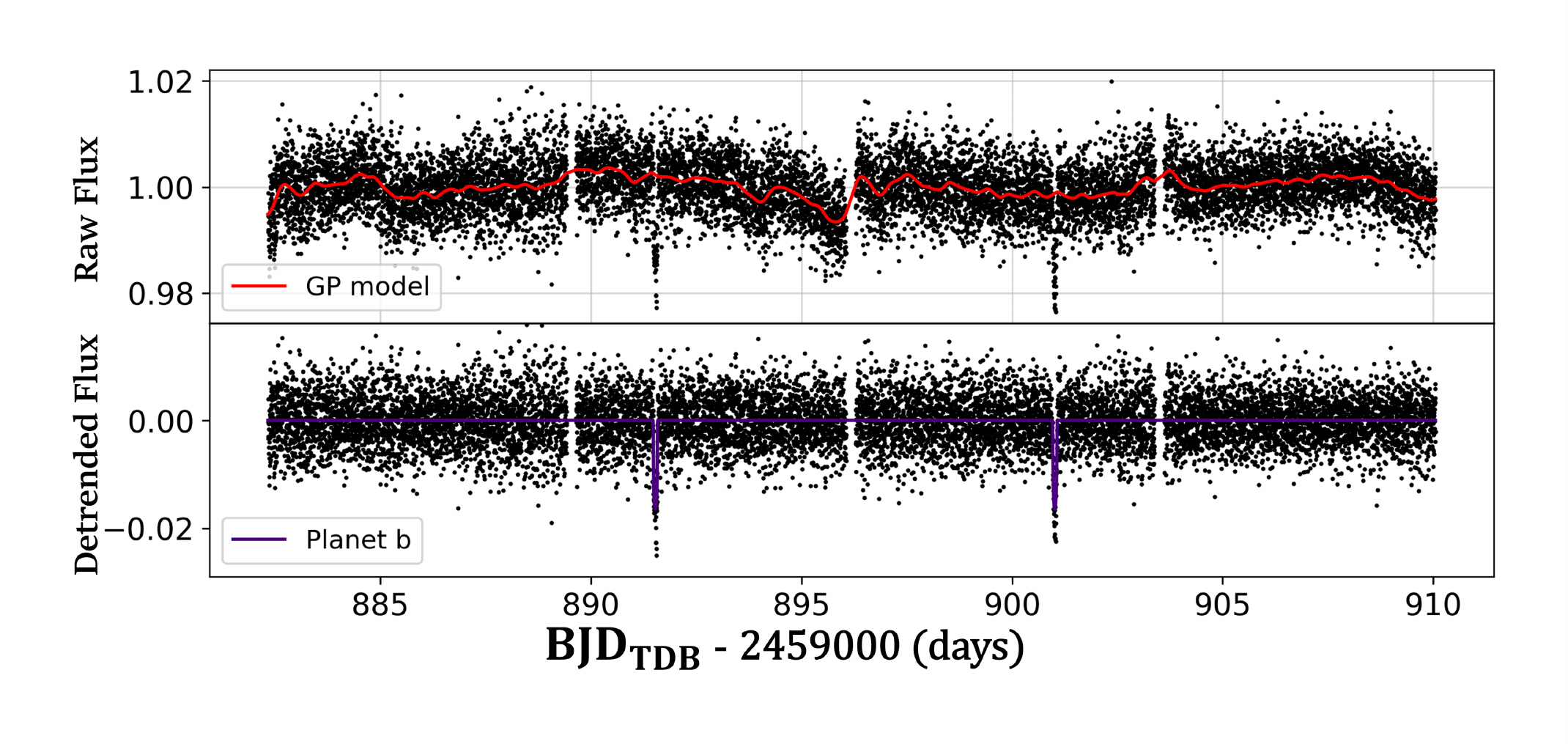}
    \includegraphics[width=\columnwidth]{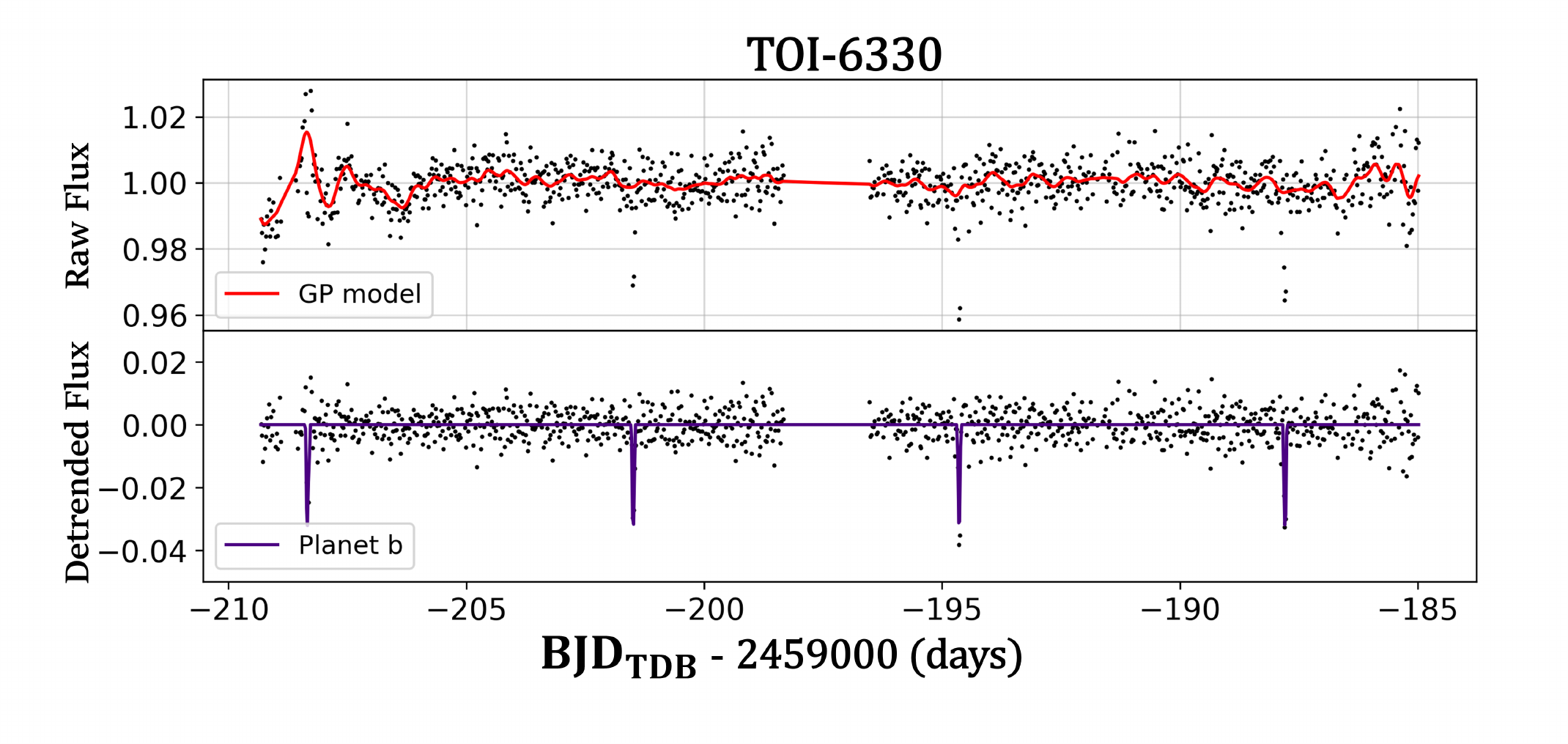}
    \includegraphics[width=\columnwidth]{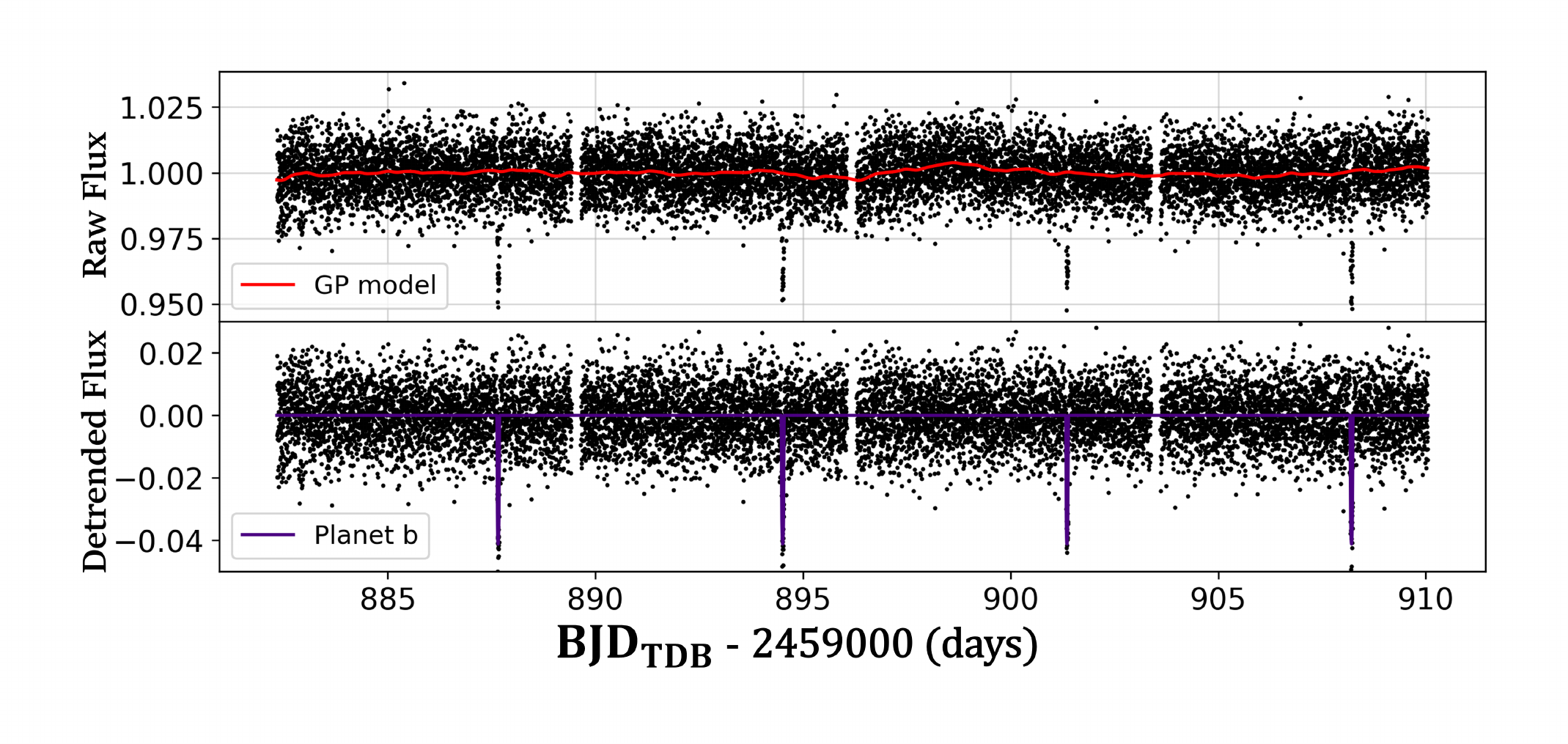}
    \caption{The TOI-6303 and TOI-6330 light curves processed using \texttt{eleanor} for TESS Sectors 18 (the top image with an 1800-second exposure time) and 58 (the bottom image with a 200-second exposure time). The top panel of each sector's light curve is a rotation GP kernel used to remove both astrophysical and instrumental systematics in the light curve labeled \texttt{RotationTerm} from \texttt{celerite2} (seen in red). The bottom panel is the best-fit GP-subtracted photometry for the transit of TOI-6303b and TOI-6330b.}
    \label{fig:6303_6330_TESS}
\end{figure}

We obtain the light curves using \texttt{eleanor} \citep{feinstein_eleanor_2019} by extracting a 31$\times$31 pixels cut-out from calibrated TESS full-frame images (FFIs). \texttt{eleanor} adopted a aperture settings of a 2$\times$1 pixel rectangle as this aperture minimized the CDPP on 1-hour long timescales and \texttt{CORR\_FLUX} values to derive the light curves from both sectors for both objects (Figure~\ref{fig:6303_6330_TESS}). 

\begin{figure*}
    \centering
    \includegraphics[width=0.95\textwidth]{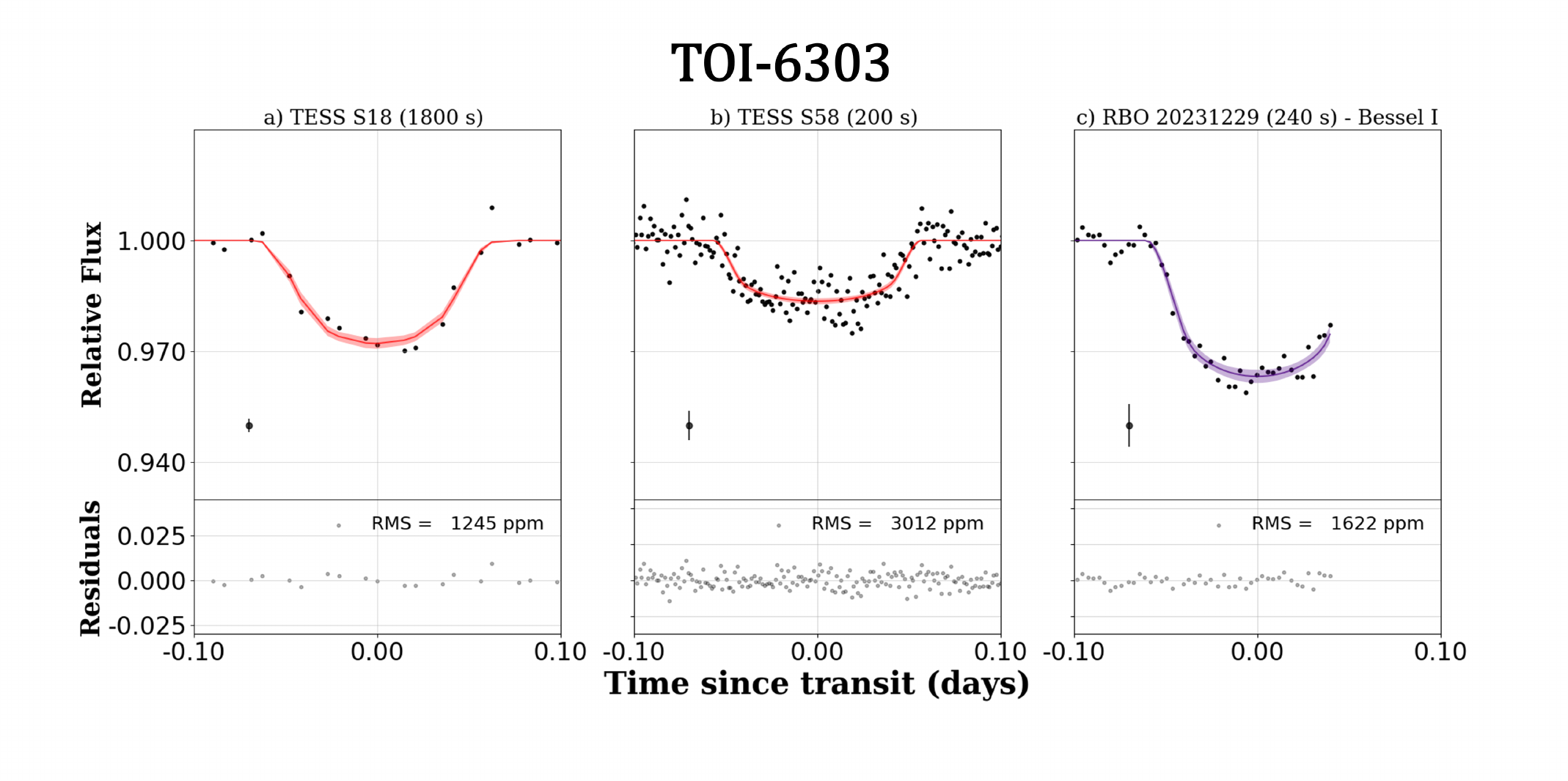}
    \includegraphics[width=0.95\textwidth]{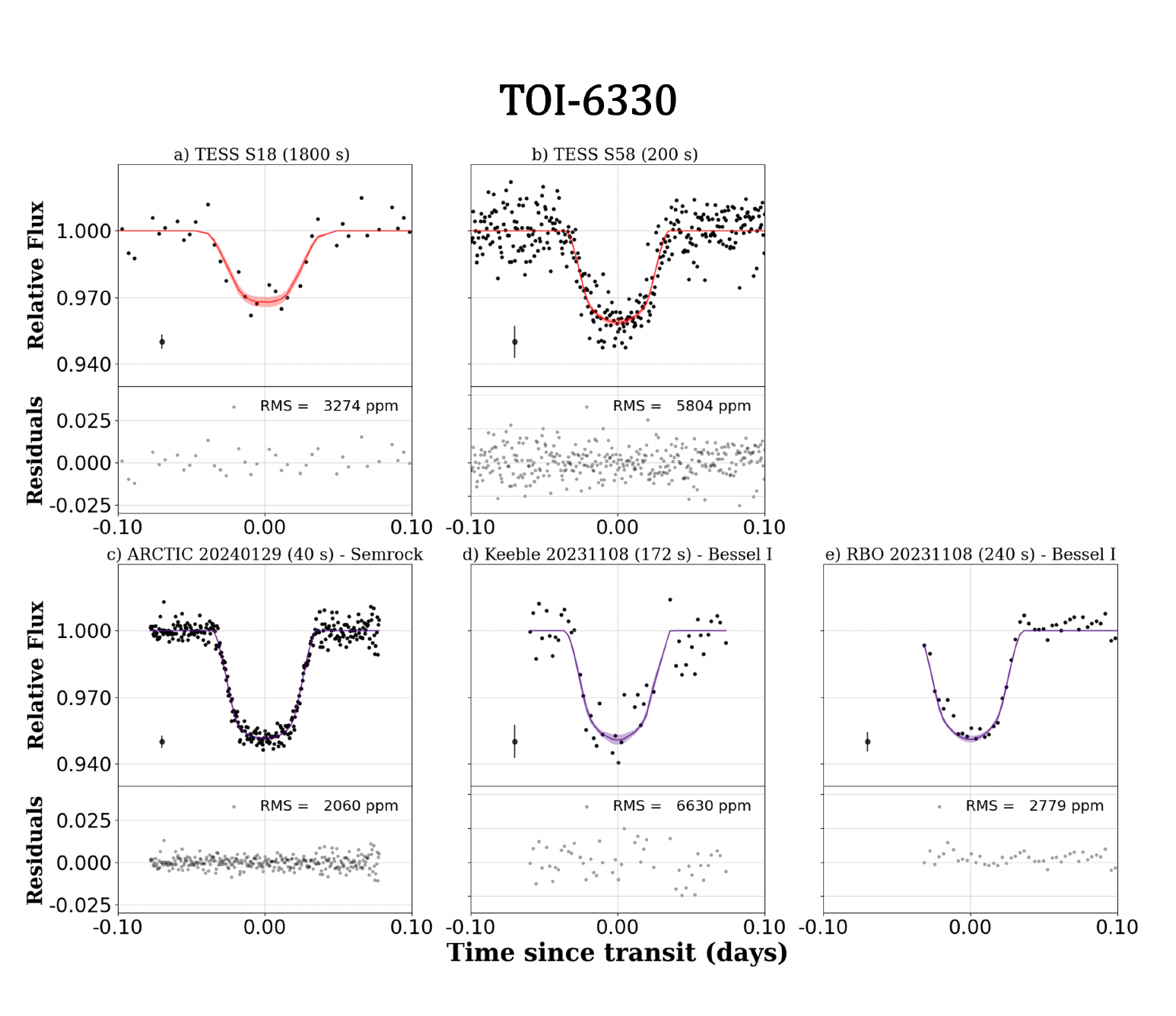}
    \caption{The TOI-6303b and TOI-6330b phase folded photometric observations. The TESS phase-folded light curves are shown in subfigures \textbf{a} and \textbf{b} for Sectors 18 and 58 respectively. The black points show the detrended data and the purple line details the model with a 1-$\sigma$ uncertainty in the translucent-purple region. We show the median uncertainty at -0.07 days. The red models for the TESS light curves indicate a floating dilution term, which is constrained with the datasets in purple.}
    \label{fig:LCPhase}
\end{figure*}

TOI-6303 and TOI-6330 both lie in a sparse region of the sky with the nearest star being $\sim$ 17$''$ and $\sim$ 10$''$ away respectively \citep{gaia_collaboration_gaia_2021}. These close by stars are still blended in the 21$''$$\times$21$''$ TESS pixel, which can be seen in the TESS light curves (see Figure~\ref{fig:LCPhase}). This dilution is corrected using ground-based photometry as discussed in section~\ref{sec:GroundBased}.

\subsection{Ground-based Photometry} \label{sec:GroundBased}
We observe one transit of TOI-6303 and three transits of TOI-6330 from the ground to confirm TOI-6303 and TOI-6330 as the host stars for the TESS-identified transit events as well as correct for undiluted transit depths. Table~\ref{tab:photosum} provides an overview of both targets' observations, telescope, and instrumental set-up.

\begin{table*}\centering
\caption{Summary of ground-based photometry.}
\begin{tabular}{ccccccc}
\hline
\hline
Object Name & Obs Date & Filter & Exposure & PSF & Field of View & Telescope \\
& (YYYY-MM-DD) &  & Time (s) & FWHM ($''$) & ($'$) \\
\hline
TOI-6330 & 2023-11-08 & Bessel I & 240 & 2.0 - 2.8 & 8.94 $\times$ 8.94 & RBO (0.6 m) \\
TOI-6330 & 2023-11-08 & Kron/Cousins I & 172 & 3.8 - 4.9 & 15.9 $\times$ 10.75 & Keeble (0.4 m) \\

TOI-6303 & 2023-12-29 & Bessel I & 240 & 1.8 - 2.7 & 8.94 $\times$ 8.94 & RBO (0.6 m) \\
TOI-6330 & 2024-01-29 & Semrock & 40 & 5.1 - 9.1 & 7.9 $\times$ 7.9 & APO (3.5 m) \\
\hline
\end{tabular}
\label{tab:photosum}
\end{table*}

\subsubsection{Red Buttes Observatory}
We observed one transit of TOI-6303b and one transit of TOI-6330b using the Red Buttes Observatory (RBO) 0.6-meter telescope in Wyoming \citep{kasper_remote_2016}.  The TOI-6303b transit was observed on the night of 2023 December 29 and the TOI-6330b transits were observed on the night of 2023 November 8 with observation details in Table~\ref{tab:photosum}. 


The observations were processed by first bias and dark subtracting each image and dividing by a median-combined normalized dome-flat. Then we looped-over randomly selected reference stars and aperture sizes to minimize the RMS scatter in the processed light curve. Further details on this processing method are described by \cite{monson_standard_2017}. The TOI-6330b light curve was further de-trended by masking out the transit and dividing out an instrumental linear trend in the baseline. The final light curves are plotted in Figure~\ref{fig:LCPhase}.

\subsubsection{Apache Point Observatory} \label{sec:APO}
We observed a transit of TOI-6330b on the night of 2024 January 29 using the Astrophysical Research Consortium (ARC) Telescope Imaging Camera \citep[ARCTIC;][]{huehnerhoff_astrophysical_2016} on the 3.5-meter telescope at Apache Point Observatory (APO). We utilized the Semrock filter \citep[842 to 873 nm][]{stefansson_toward_2017} with an exposure time of 40 seconds and airmass increased from 1.6 to 5.7. The Semrock filter bandpass minimizes overlap with atmospheric water-band features reducing photometric scatter due to varying airmass or atmospheric water columns. The exposures were captured utilizing 4$\times$4 binning. We used the fast readout mode which allowed us to obtain a readout time of 1.3 seconds, a gain of 2 e$^-$/ADU, and a plate scale of 0.456$''$/pixel.

We bias subtracted and flat fielded the data using \texttt{AstroImageJ} \citep{collins_astroimagej_2017}. We performed differential photometry with an aperture radius of 8 pixels (3.65$''$) and an inner sky radius of 14 pixels (6.38$''$) alongside an outer sky radius of 20 pixels (9.12$''$) used to subtract the star's background. We calculated the flux uncertainties using a combination of photon noise, detector read noise, and airmass \citep{stefansson_toward_2017}. The light curve is plotted on the bottom right of Figure~\ref{fig:LCPhase}. 



\subsubsection{Keeble Observatory}
We observed a transit of TOI-6330b on 2023 November 8 using the 0.4-meter telescope at the Randolph-Macon College’s Keeble Observatory in Ashland, Virginia. We used the Kron/Cousins I-Band filter with an exposure time of 172 seconds. The observations were obtained using 2$\times$2 binning with an airmass decreasing from 1.3 to 1.0. We made our observations with a plate scale of 0.88$''$ and a gain of 1.3e$^-$/ADU. The data was reduced and analyzed using \texttt{AstroImageJ} \citep{collins_astroimagej_2017} following the same procedure as APO (Section~\ref{sec:APO}). We used an aperture radius of 8 pixels (7.032$''$), an inner sky radius of 15 pixels (13.185$''$), and an outer sky radius of 20 pixels (17.58$''$). See the bottom middle panel in Figure~\ref{fig:LCPhase} for the light curve. 

\subsection{Radial Velocity}\label{RVSection}

\subsubsection{HPF}
We obtained radial velocity (RV) observations for TOI-6303 and TOI-6330 using the Habitable-zone Planet Finder \citep[HPF;][]{mahadevan_habitable-zone_2012,mahadevan_habitable-zone_2014} located at the 10-meter Hobby Eberly Telescope \citep[HET][]{ramsey_early_1998}. HPF is a near-infrared, environmentally stabilized \citep{stefansson_versatile_2016}, fiber-fed \citep{kanodia_overview_2018}, high-resolution precision spectrograph with a resolution of 55,000 spanning from 810 -- 1280 nm. HPF is fully queue-scheduled with HET resident astronomers executing all observations \citep{shetrone_ten_2007}.

TOI-6303 was observed seven times between 2023 July 29 to 2023 November 6 with an exposure time of 945 seconds. Four of the seven observations consist of two subsequent exposures which were later binned post-processing to minimize noise. TOI-6330 was observed nine times between 2023 August 30 to 2023 September 28, with each visit consisting of two exposures of 945 seconds each. The RVs from individual exposures were combined by weighted averaging for each night.  

The data was processed using the \texttt{HxRGproc} package \citep{ninan_habitable-zone_2018}. The barycentric corrections were performed with \texttt{barycorrpy} \citep{kanodia_python_2018}, which is the Python implementation of the algorithms from \cite{wright_barycentric_2014}. We did not use the near-infrared (NIR) Laser Frequency Comb \citep{metcalf_stellar_2019} for HPF due to concerns about scattered light. 



\begin{figure*}[!t]
    \centering
    \includegraphics[width=0.65\textwidth]{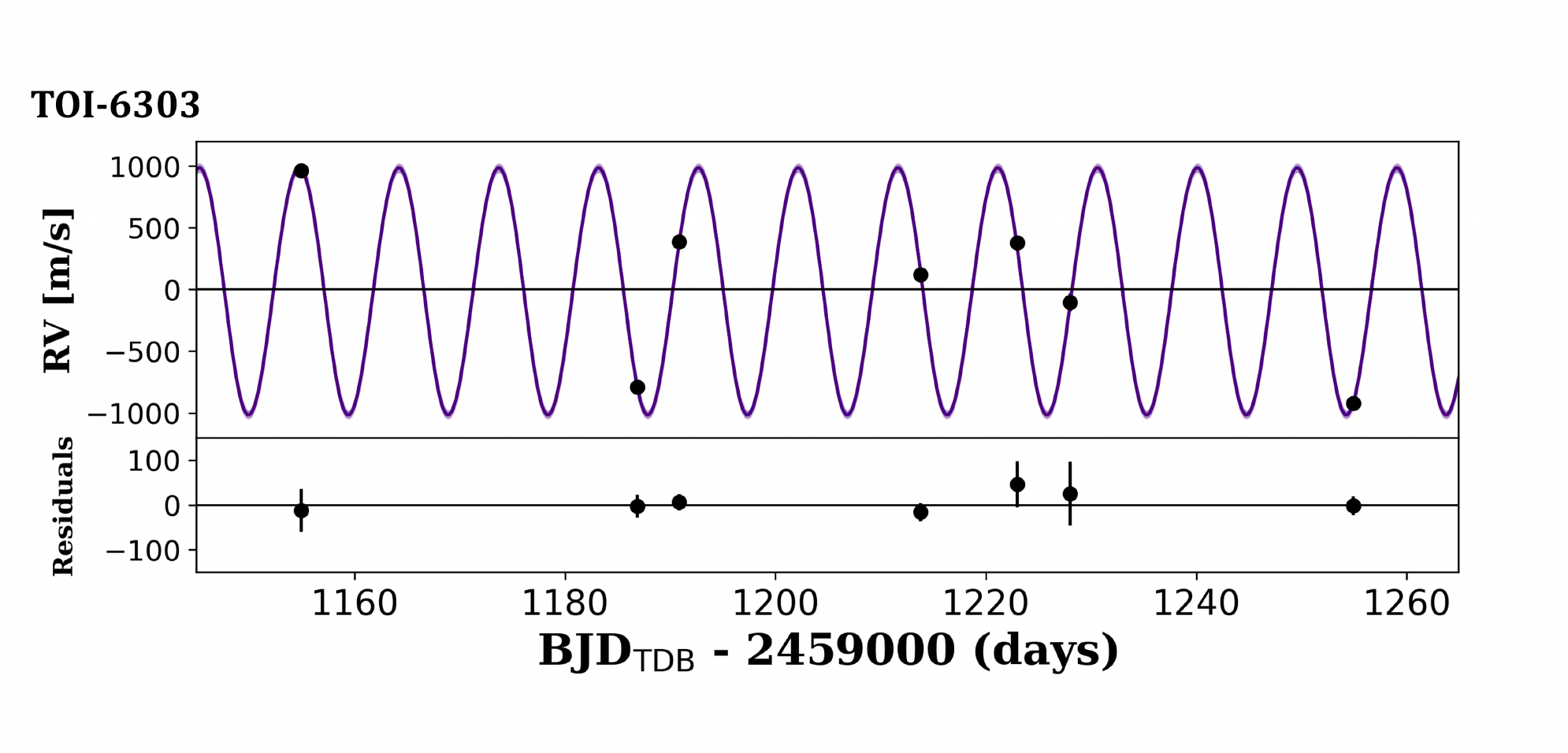}\hfill
    \includegraphics[width=0.35\textwidth]{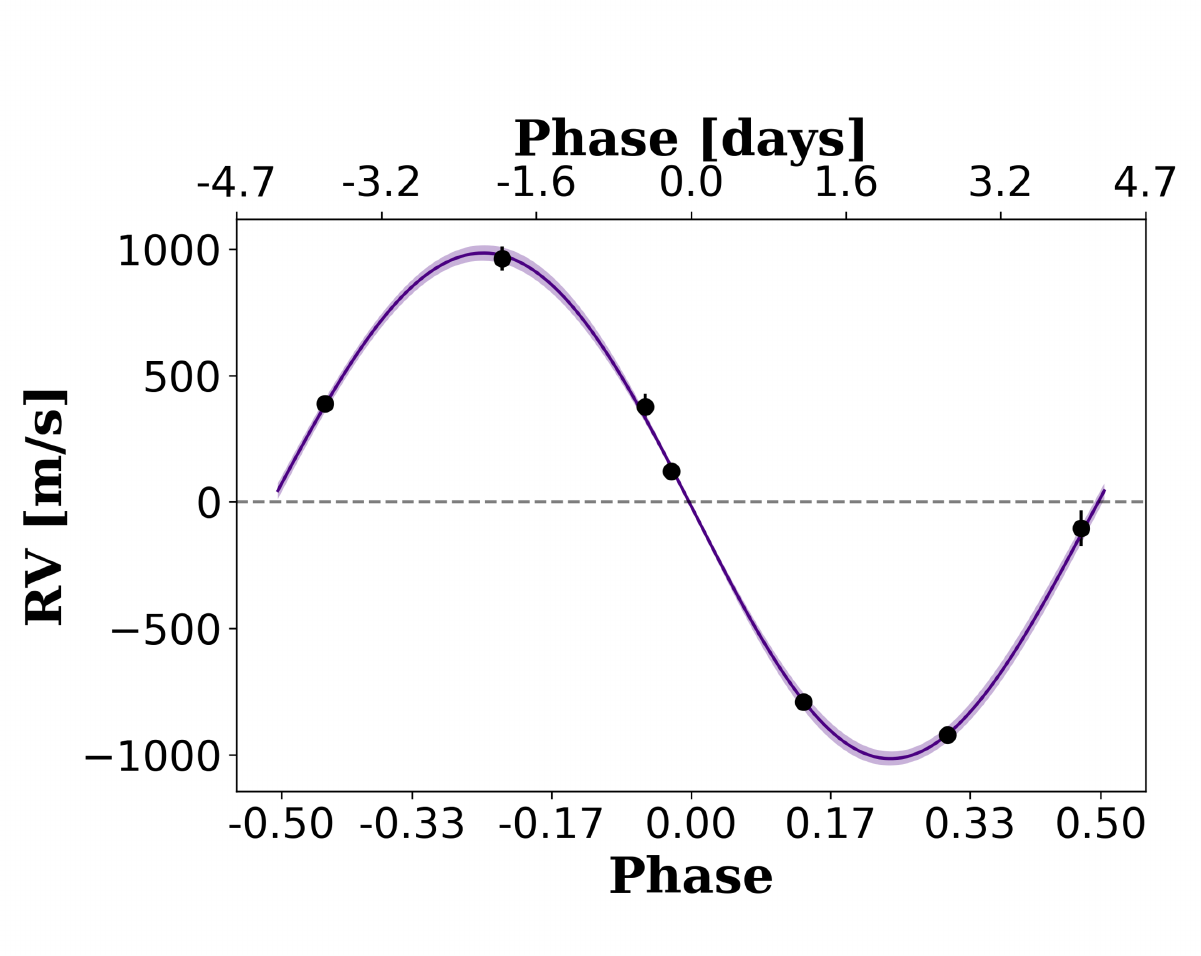}
    \includegraphics[width=0.65\textwidth]{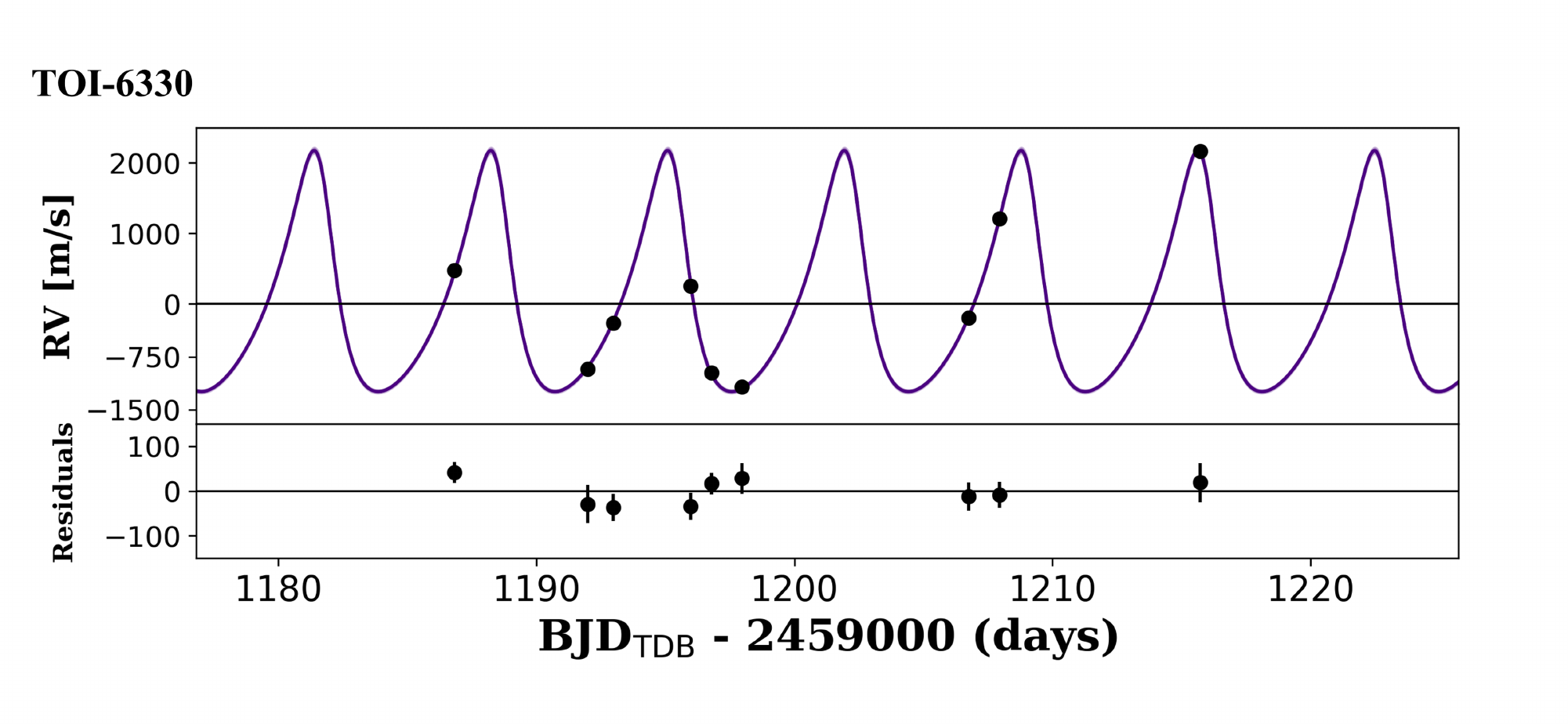}\hfill
    \includegraphics[width=0.35\textwidth]{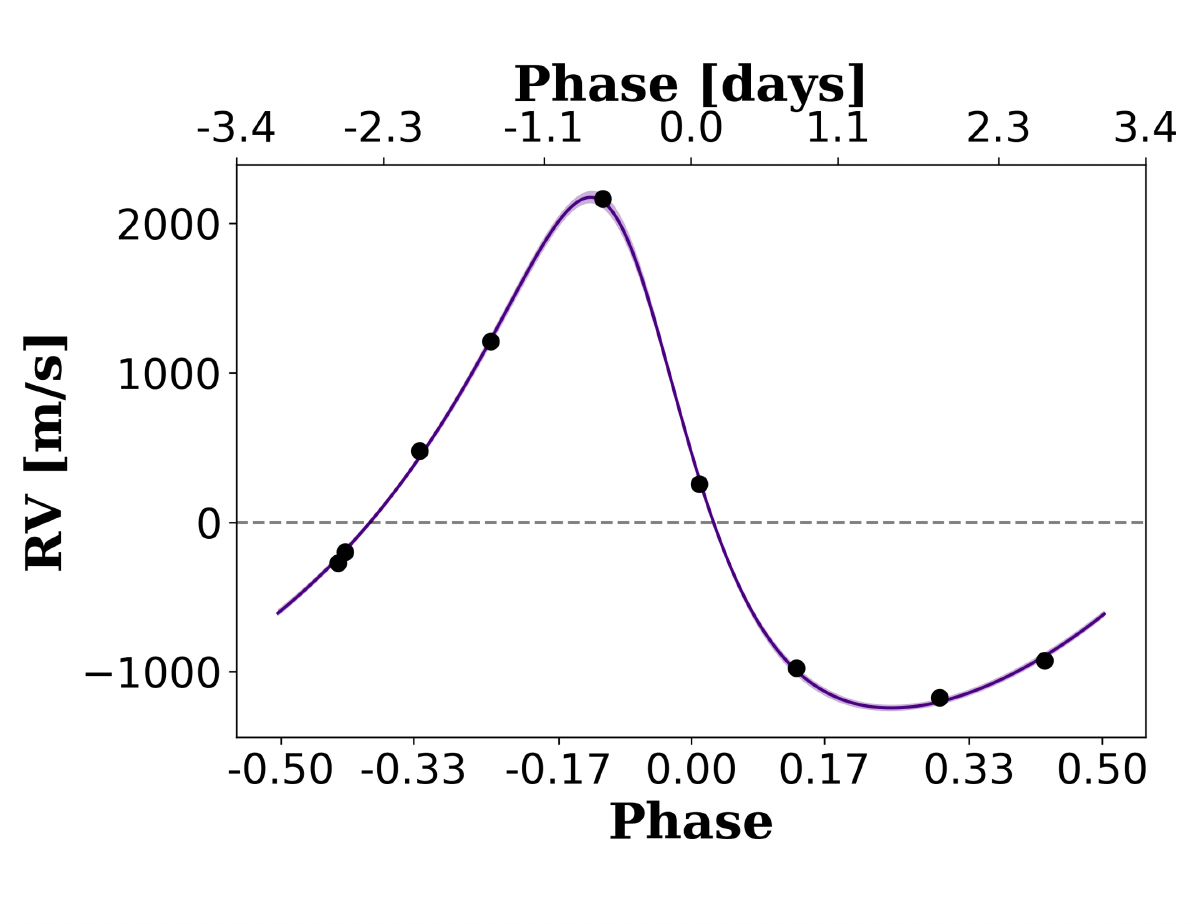}
    \caption{\textbf{Left}: A RV time series for TOI-6303 (\textbf{Top}) and TOI-6330 (\textbf{Bottom}) as observed by HPF. The model best fit is designated by the purple line with a lighter purple indicating the 16-84\% uncertainty. \textit{Note: For TOI-6303, the observations with larger error bars are one 15-minute exposure while the rest are two binned 15-minute exposures (See Table~\ref{tab:RVs}). } \textbf{Right}: Phase-folded HPF RVs on best-fit orbital parameters (\textbf{Top:} TOI-6303, \textbf{Bottom:} TOI-6330). We set the eccentricity and arugment of periastron to float. TOI-6303 is consistent with a circular orbit while TOI-6330 shows an eccentric orbit.}
    \label{fig:RVSeriesPhase}
\end{figure*}

We applied a version of the template-matching algorithm \texttt{SERVAL} \citep{zechmeister_spectrum_2018, stefansson_neptune-mass_2023}, modified for HPF \citep{metcalf_stellar_2019} to estimate the RVs from the spectra for both systems. Using this method, we first create a master template of all observations and then compare the Doppler shift of each spectra to the master template to minimize $\chi^2$ statistics. The master template is constructed after the masking of the telluric and sky-emission lines. Our final binned RVs for TOI-6303 and TOI-6330 are listed in Table~\ref{tab:RVs} and shown in Figure~\ref{fig:RVSeriesPhase}

\begin{table}[]\centering
\begin{tabular}{ccc}
\hline
\hline
Date (BJD$_{\mathrm{TDB}}$) & RV (m s$^{-1}$) & $\sigma_{\mathrm{RV}}$ (m s$^{-1}$) \\ 
\hline
& TOI-6303 \\
\hline
2460154.94681$^{a}$               & -967          & 48                                \\
2460186.86262               & -785           & 25                                \\
2460190.85569               & 393           & 18                                \\
2460213.80174               & 126          & 20                                \\
2460222.98768$^{a}$               & 382          & 51                                \\
2460227.98737$^{a}$               & -100           & 71                                \\
2460254.91193               & -915           & 21                                \\
\hline
& TOI-6330 \\
\hline
2460186.81214               & 271          & 24                                \\
2460191.98574               & -1132           & 42                                \\
2460192.98611               & -481           & 30                                \\
2460195.97687               & 48          & 30                                \\
2460196.78045               & -1185          & 24                                \\
2460197.96543               & -1381           & 34                                \\
2460206.74331               & -408           & 31                                \\
2460207.94776               & 1004           & 29                                \\
2460215.72917               & 1958           & 44                                \\
\hline
\end{tabular}
\caption{The 30-minute weighted-average binned HPF RVs of TOI-6303 and TOI-6330. \textit{Note: Observations denoted with an $^{a}$ have a single 15-minute exposure}}
\label{tab:RVs}
\end{table}

\subsection{Speckle Imaging}

\begin{figure}
    \centering
    \includegraphics[width=0.5\textwidth]{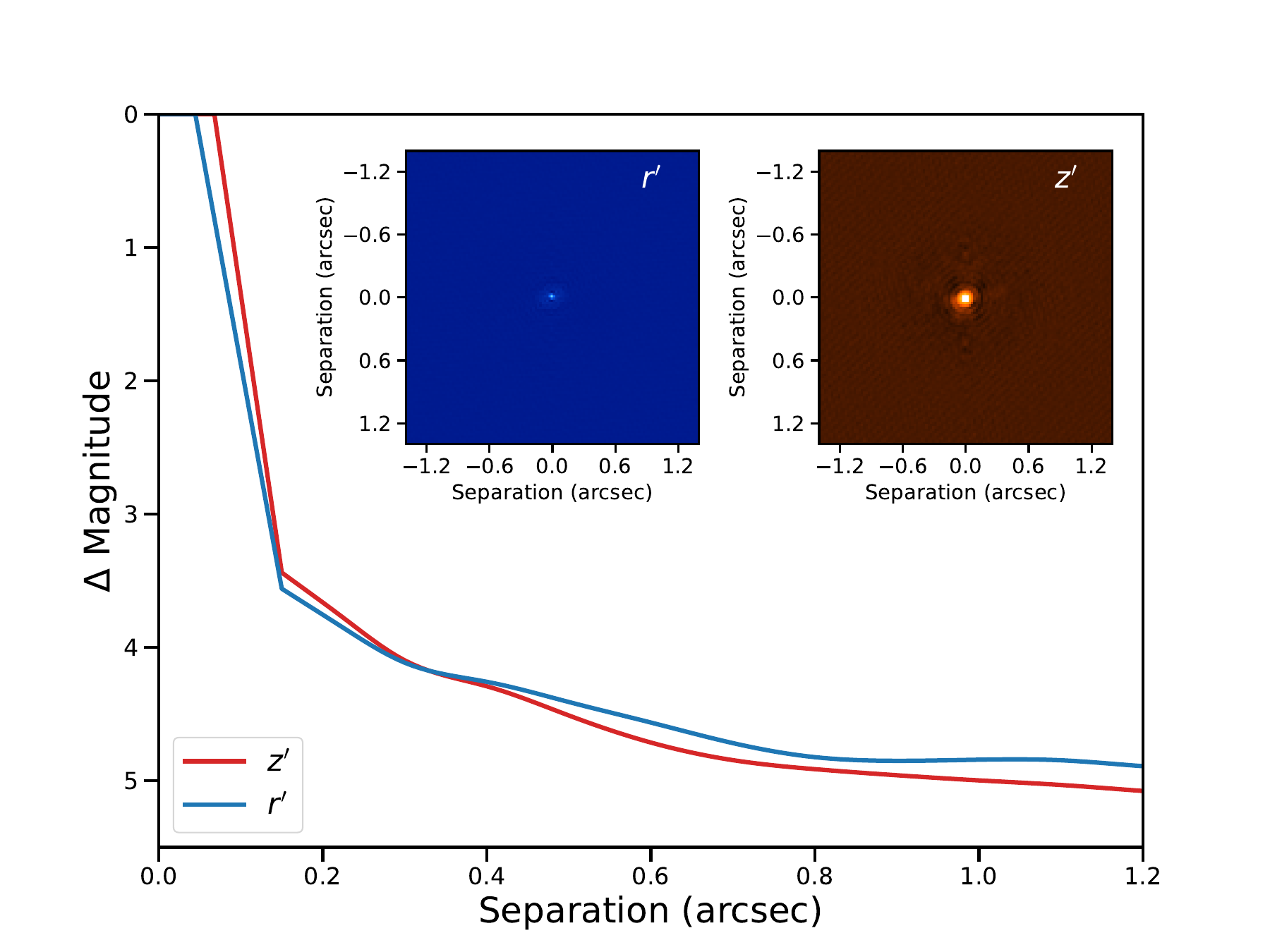}\hfill
    \includegraphics[width=0.5\textwidth]{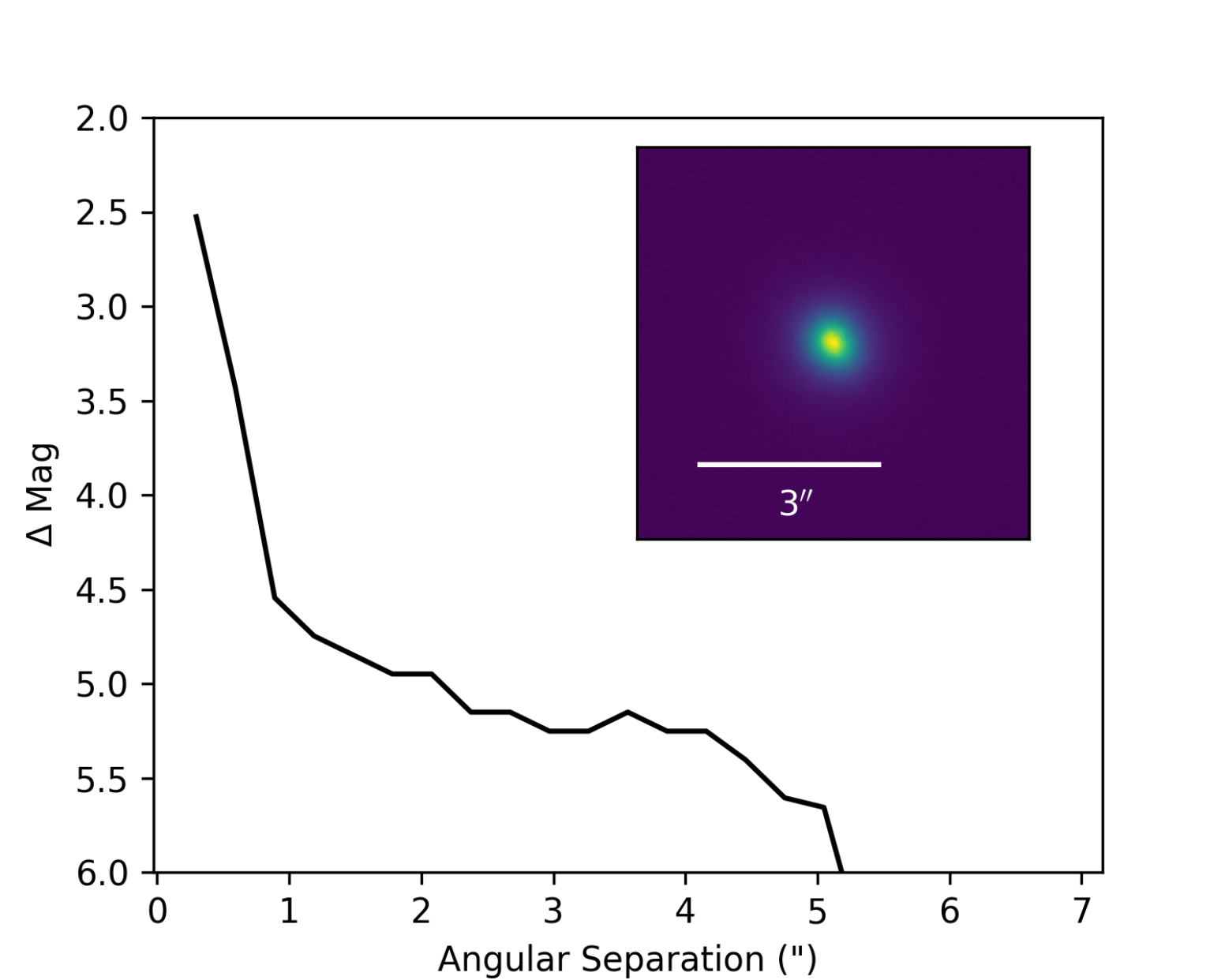}
    \caption{\textbf{Top:} 5$\sigma$ processed contrast curve for TOI-6303 observed from NESSI in the Sloan z' and r' filters. \textbf{Bottom:} 5$\sigma$ processed contrast curve for TOI-6303 observed from ShaneAO in the Ks-band.}
    \label{fig:SpecklePlots6303}
\end{figure}

To rule out nearby background sources or faint stellar companions, we obtained data from the NN-EXPLORE Exoplanet Stellar Speckle Imager \citep[NESSI;][]{scott_nn-explore_2018} located on the WIYN\footnote{The WIYN Observatory is a joint facility of the NSF’s National
Optical-Infrared Astronomy Research Laboratory, Indiana Uni-
versity, the University of Wisconsin-Madison, Pennsylvania State
University, Purdue University and Princeton University.} 3.5-meter telescope at Kitt Peak National Observatory and the ShARCS camera \citep{srinath_swimming_2014} located on the Shane 3-meter telescope at the University of California's Lick Observatory. 

TOI-6303 observations occurred on 2024 February 17 and 2023 November 25 with NESSI and ShaneAO respectively. The NESSI observations were a 9-minute sequence of 40 ms exposures for both systems using a Sloan r' and z' filter. We processed and combined the exposures using the method described by \cite{howell_speckle_2011}. The ShaneAO observations were taken in Laser Guide Star (LGS) mode with a 4-position dithering pattern for a total of 25-minute exposure time in Ks filter.  We detected no background sources with a separation $>$1.0$''$ and $\Delta Ks$ $<$ 4.5 for ShaneAO, and no background sources with a $>$0.3$''$ and $\Delta z'$, $\Delta r'$ $<$ 4.0 for NESSI.

\begin{figure}
    \centering
    \includegraphics[width=0.5\textwidth]{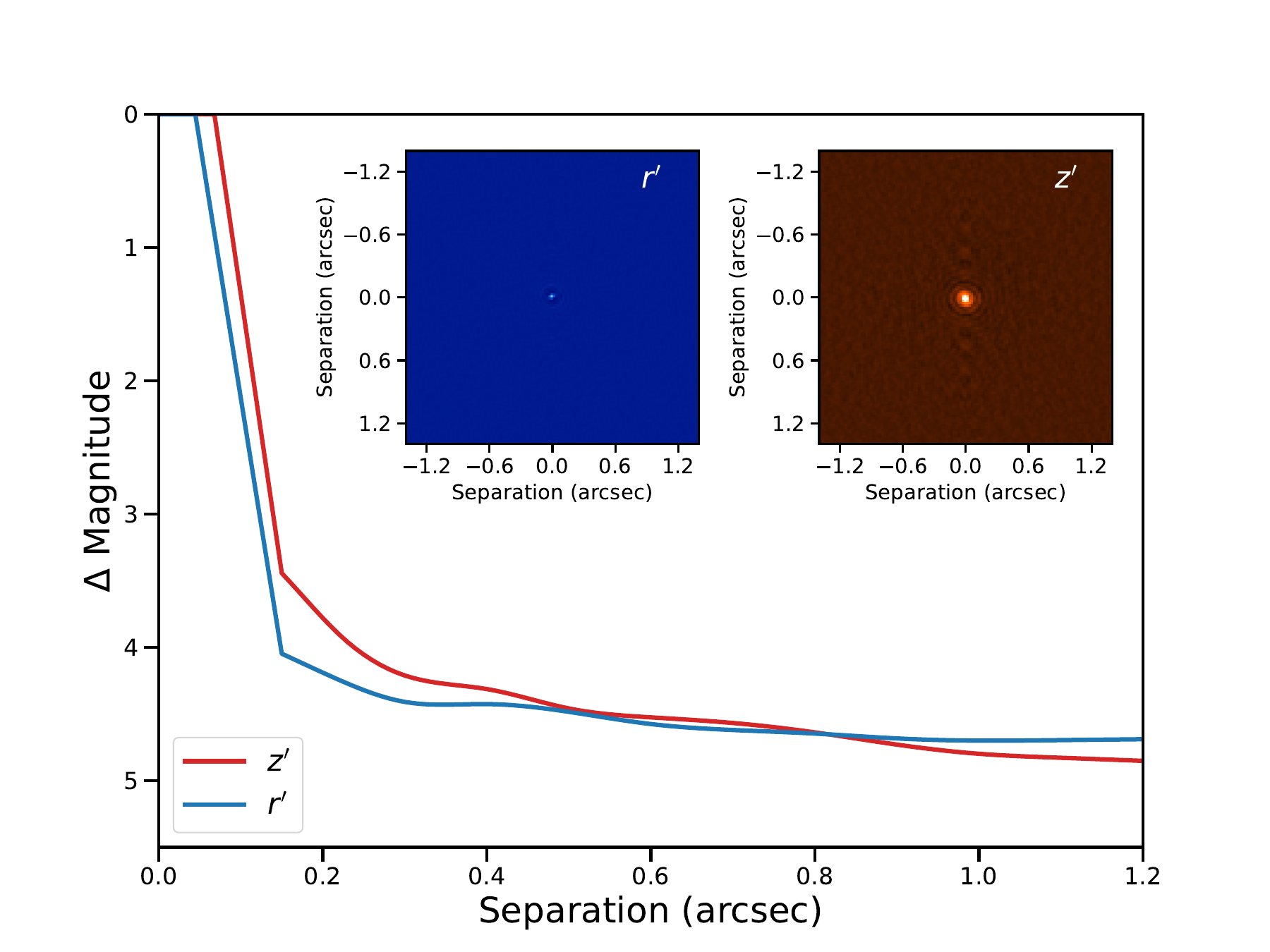}
    \includegraphics[width=0.5\textwidth]{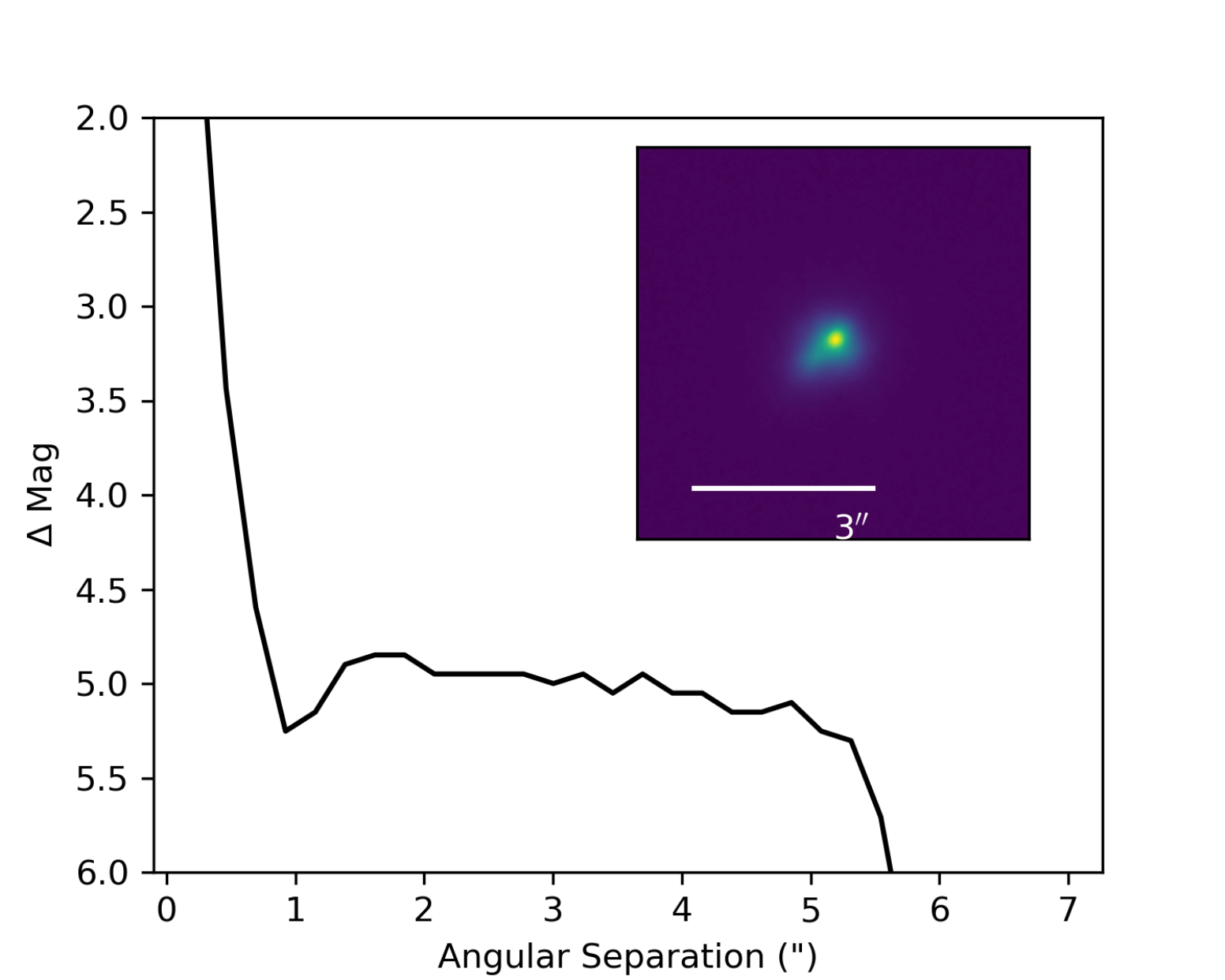}
    \caption{\textbf{Top:} 5$\sigma$ processed contrast curve for TOI-6330 observed from NESSI in the Sloan z' and r' filters. \textbf{Bottom:} 5$\sigma$ processed contrast curve for TOI-6330 observed from ShaneAO in the Ks-band.}
    \label{fig:SpecklePlot6330}
\end{figure}

TOI-6330 was observed on 2023 September 8 with NESSI with the same observation and post-processing techniques as for the TOI-6303 observations. TOI-6330 was also observed on 2024 July 25 by ShaneAO in Non-Laser Guide Star mode with 5-position dithering for a total of 32-minute exposure time in Ks filter. As for TOI-6303, We detected no background sources with a separation $>$1.0$''$ and $\Delta Ks$ $<$ 5.25 for ShaneAO and no background sources with a separation $\>$0.3$''$ and $\Delta z'$, $\Delta r'$ $<$ 4.0. TOI-6303 and TOI-6330 contrast curves are seen in Figure~\ref{fig:SpecklePlots6303} and ~\ref{fig:SpecklePlot6330}.  


\section{Analysis}\label{sec:Analysis}
\subsection{Stellar Parameters}
The stellar parameters for both systems were determined using the \texttt{HPF-SpecMatch} package \citep{stefansson_sub-neptune-sized_2020} for spectroscopic parameters. The fits were performed by matching the target spectra against a catalog of HPF spectra for GKM dwarfs to calculate a $\chi^2$ value. This $\chi^2$ value is then used to determine the five best catalog stars utilized to characterize TOI-6303 and TOI-6330. We estimate values for the effective temperatures $T_{\text{eff}}$, surface gravity $\text{log}{g}$, metallicities [Fe/H], and projected rotational velocities $v$ \text{sin} $i_{*}$ for both systems (see Table~\ref{tab:stellarparam}). Due to the minimal telluric line contamination, we use HPF's Order 5 (8534 - 8645 \AA) to fit our systems. HPF's instrumental resolution limits our $v$ sin $i_*$ to $<$ 2 km/s. Determining the metallicity for M-dwarfs is exceptionally difficult \citep{passegger_metallicities_2022} yielding a recommendation to interpret the metallicities of these stars with caution. We therefore suggest a categorical classification for TOI-6303 and TOI-6330 as super-solar and solar metallicities respectively. 

We use the \texttt{ExoFASTv2} package \citep{eastman_exofastv2_2019} to perform an SED and isochrones fit for the stellar physical parameters. The Gaussian priors for the SED fit we used were the spectroscopic parameters derived from \texttt{HPF-SpecMatch}, the distance calculated from \cite{bailer-jones_estimating_2021} using the GAIA DR3 parallax, and the magnitudes listed in Table~\ref{tab:stellarparam} \citep{cutri_2mass_2003, wright_wise_2010,magnier_pan-starrs_2020}. The stellar mass and radius for TOI-6303 and TOI-6330 are 0.644 $\pm$ 0.024 M$_\odot$, 0.609 $\pm$ 0.016 R$_\odot$ and 0.531 $\pm$ 0.021 M$_\odot$, 0.490 $\pm$ 0.011 R$_\odot$ respectively. Physical parameters for both stars are detailed in Table~\ref{tab:stellarparam}. 

\begin{figure}[!t]
    \centering
    \includegraphics[width=0.5\textwidth]{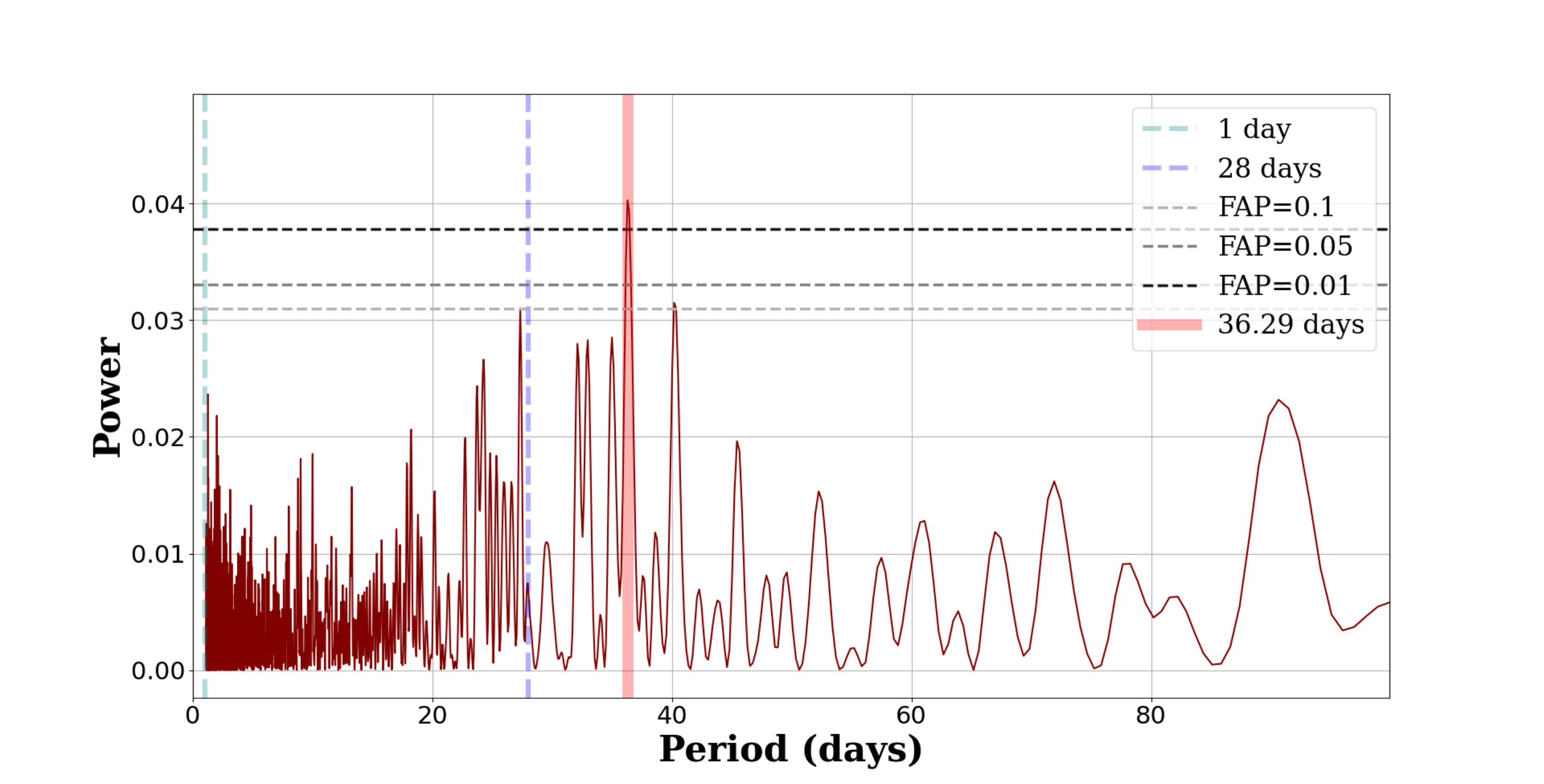}
    \caption{The TOI-6330 ZTF periodogram with the ZTF-r filter. The teal and blue vertical lines detail the 1 and 28-day period respectively. The red highlight shows a peak period at 36.29 days believed to be from stellar activity. The horizontal dashed lines detail the 10\%, 5\%, and 1\% FAPs from bottom to top. The ZTF data is obtained directly from the ZTF database via IRSA}
    \label{fig:6330_ZTF}
\end{figure}

We utilized ground-based monitoring with the Zwicky Transient Facility \citep[ZTF;][]{masci.ztf} alongside TESS to search for rotational variability in the photometry. We do not detect a significant peak in the ZTF periodogram for TOI-6303 in both the ZTF-r and ZTF-g filters. From the HPF limit of $v$ sin $i_*$ $<$ 2 km/s, TOI-6303 has a lower limit rotation period of $\sim$ 15 days assuming a near 90-degree inclination, however, given the non-detection, the rotation period is likely longer than this minimum. We detect a peak above the 1\% False Alarm Probability (FAP) line in the ZTF-r periodogram at an approximately 36 day rotation period for TOI-6330 (Figure~\ref{fig:6330_ZTF}). We also utilized the ZTF-g periodogram to attempt to detect any activity but there is no peak greater than the 10\% FAP line. We believe this non-detection in the g-band is due to the faintness of the star alongside the possible interference of the lunar cycle. Furthermore, we do not detect any periodicity in the activity indicators measured by HPF-\texttt{SERVAL}. It is important to note with only nine binned observations, it is challenging to detect a 36-day rotation period with any confidence. The HPF limit of $v$ sin $i_*$ $<$ 2 km/s yields a lower limit rotation period of $\sim$ 12 days assuming a near 90-degree inclination. We, therefore, cautiously suggest a 36-day rotation period for TOI-6330. 

We use the Renormalized Unit Weight Error (RUWE) from Gaia DR3 \citep{gaia_collaboration_gaia_2021} to determine the likelihood of a blended companion leading to excess astrometric noise. A RUWE value of $\sim$1 is consistent with a single star while a value $>$1.4 is a conservative threshold indicating a massive companion \citep{ziegler_ruwe_2020, gaia.ruwe}. TOI-6303 and TOI-6330 have RUWE values of 1.01 and 1.05 respectively indicating a single source with no additional light blending into the transits from other stars for both systems.

The regions in which TOI-6303 and TOI-6330 are located are sparse regions of the sky and none of the resolved companions in Gaia DR3 within 60$''$ have similar proper motion or parallax. 

\begin{deluxetable*}{lcccc}
\tablecaption{Summary of stellar parameters. \label{tab:stellarparam}}
\tablehead{\colhead{~~~Parameter}&  \colhead{Description}&
\colhead{TOI-6303}&
\colhead{TOI-6330}&
\colhead{Reference}}
\startdata
\multicolumn{5}{l}{\hspace{-0.2cm} Main identifiers:}  \\
~~~TIC & \tess{} Input Catalogue  & TIC-186810676 & TIC-308120029 & Stassun \\
~~~2MASS & \(\cdots\) & 2MASS J03070731+4008503 & 2MASS J01284037+5341105 & 2MASS  \\
~~~Gaia DR3 & \(\cdots\) & Gaia DR3 239050153051494272 & Gaia DR3 407530931116600320 & Gaia DR3\\
\multicolumn{5}{l}{\hspace{-0.2cm} Equatorial Coordinates, Proper Motion and Distance:} \\
~~~$\alpha_{\mathrm{J2000}}$ &  Right Ascension (RA) & 03:07:07.36 & 01:28:40.32 & Gaia DR3\\
~~~$\delta_{\mathrm{J2000}}$ &  Declination (Dec) & +40:08:49.83 & +53:41:09.88 & Gaia DR3\\
~~~$\mu_{\alpha}$ &  Proper motion (RA, \unit{mas/yr}) & 32.941 $\pm$ 0.069 & -35.668 $\pm$ 0.164 & Gaia DR3\\
~~~$\mu_{\delta}$ &  Proper motion (Dec, \unit{mas/yr}) & -35.375 $\pm$ 0.054 & -45.0748 $\pm$ 0.073 & Gaia DR3 \\
~~~$\omega$ &  Parallax (mas)  & 6.587 $\pm$ 0.019 & 6.967 $\pm$ 0.027 & Gaia DR3 \\
~~~$d$ &  Distance in pc  & 151.82 $\pm$ 0.42 & 143.51 $\pm$ 0.55 & Bailer-Jones \\
~~~\(A_{V,max}\) & Maximum visual extinction & 0.1 & 0.1 & Green\\
\multicolumn{5}{l}{\hspace{-0.2cm} Optical and near-infrared magnitudes:}  \\
~~~$V$ & Johnson V mag & $14.6 \pm 0.2$ & $16.3 \pm 0.2$ & APASS\\
~~~$G$ & Mean G mag & 13.8979 $\pm$ 0.0004 & 14.9120 $\pm$ 0.0005 & Gaia DR3\\
~~~$g^{\prime}$ &  Pan-STARRS1 $g^{\prime}$ mag  & 15.885 $\pm$ 0.0004 & 16.492 $\pm$ 0.0071 & Pan-STARRS1\\
~~~$r^{\prime}$ &  Pan-STARRS1 $r^{\prime}$ mag  & 14.194 $\pm$ 0.002 & 15.317 $\pm$ 0.0026 & Pan-STARRS1 \\
~~~$i^{\prime}$ &  Pan-STARRS1 $i^{\prime}$ mag  & 13.607 $\pm$ 0.003 & 14.260 $\pm$ 0.0032 & Pan-STARRS1 \\
~~~$z^{\prime}$ &  Pan-STARRS1 $z^{\prime}$ mag  & 13.378 $\pm$ 0.005 & 13.772 $\pm$ 0.0016 & Pan-STARRS1 \\
~~~$J$ & $J$ mag & 11.679 $\pm$ 0.018 & 12.368 $\pm$ 0.023 &2MASS\\
~~~$H$ & $H$ mag & 11.036 $\pm$ 0.017 & 11.744 $\pm$ 0.015 &2MASS\\
~~~$K_s$ & $K_s$ mag & 10.826 $\pm$ 0.016 & 11.513 $\pm$ 0.023 &2MASS\\
~~~$W1$ &  WISE1 mag & 10.747 $\pm$ 0.023 & 11.356 $\pm$ 0.021 &WISE\\
~~~$W2$ &  WISE2 mag & 10.786 $\pm$ 0.026 & 11.299 $\pm$ 0.020 &WISE\\
~~~$W3$ &  WISE3 mag & 10.25 $\pm$ 0.067 & 11.192 $\pm$ 0.136 &WISE\\   
\multicolumn{5}{l}{\hspace{-0.2cm} \texttt{SpecMatch} Spectroscopic Parameters:}\\
~~~$T_{\mathrm{eff}}$ &  Effective temperature in \unit{K} & 3977 $\pm$ 59 & 3539 $\pm$ 59 & This work\\
~~~$\mathrm{[Fe/H]}$ &  Metallicity in dex & 0.44 $\pm$ 0.16 & 0.08 $\pm$ 0.16 &This work\\
~~~$\log(g)$ & Surface gravity in cgs units & 4.67 $\pm$ 0.04 & 4.77 $\pm$ 0.04 &This work\\
~~~$v \sin i_*$ &  Rotational velocity in \unit{km/s}  & $<$ 2 & $<$ 2 &This work\\
~~~$P_{rot}$ &  Rotation Period in \unit{days}  & \(\cdots\) & $\sim$ 36 &This work\\
\multicolumn{5}{l}{\hspace{-0.2cm} Model-Dependent Stellar SED and Isochrone Fit Parameters:}\\
~~~$M_*$ &  Mass in $M_{\odot}$ & 0.644	$\pm$ 0.024 & 0.531 $\pm$ 0.021 &This work\\
~~~$R_*$ &  Radius in $R_{\odot}$ & 0.609 $\pm$ 0.016 & 0.490 $\pm$ 0.011 &This work\\
~~~$L_*$ &  Luminosity in $L_{\odot}$ & 0.0819 $\pm$ 0.0018 & 0.03701 $\pm$ 0.0009 &This work\\
~~~$\rho_*$ &  Density in $\unit{g/cm^{3}}$ & 4.01 $\pm$ 0.28 & 6.02 $\pm$ 0.46 &This work\\
~~~Age & Age in Gyrs & 6.4$^{+4.7}_{-4.4}$ & 7.6$^{+4.2}_{-4.9}$ &This work\\
\enddata
\tablenotetext{}{References are: Stassun \citep{stassun_tess_2018}, 2MASS \citep{cutri_2mass_2003}, Gaia DR3 \citep{gaia_collaboration_gaia_2021}, Bailer-Jones \citep{bailer-jones_estimating_2021}, Green \citep{green_3d_2019}, APASS \citep{henden_apass_2018}, Pan-STARRS1 \citep{chambers_pan-starrs1_2016,magnier_pan-starrs_2020}, WISE \citep{wright_wise_2010}}
\end{deluxetable*}

\subsection{Joint-Fit of RVs and Photometry - Planetary Parameters}

We performed a joint fit of all transit data from TESS and ground-based photometry alongside HPF RVs using the \texttt{exoplanet} package \citep{foreman-mackey_exoplanet_2021} for both systems. We detrended the TESS photometry using a Gaussian Process (GP) \texttt{RotationTerm} kernel before the joint fit while masking out the transiting events \citep{foreman-mackey_fast_2017, foreman-mackey_scalable_2018}. We fit each instrument with a quadratic limb darkening prior and a white noise jitter term (TESS Sectors + ground-based). The TESS sector pixels are large enough ($>$ 21$''$) to potentially cause contamination from neighboring stars, which we account for by including a dilution term for each TESS sector, that is estimated using the ground-based photometric data. Equation 1 in \cite{kanodia_toi-5205_2023} is used to account for dilution. We allow eccentricity and argument of periastron to float for both systems and include additional terms for RV jitter, RV offset, and a linear RV trend.

To obtain the joint-fit parameters, we employ the \texttt{exoplanet} package \citep{foreman-mackey_exoplanet_2021} using \texttt{pymc3} \citep{salvatier_probabilistic_2016} and \texttt{theano} \citep{the_theano_development_team_theano_2016} to perform MCMC sampling for both systems. We include the stellar mass (0.644 $\pm$ 0.024 M$_{\odot}$, 0.531 $\pm$ 0.021 M$_{\odot}$), stellar radius (0.609 $\pm$ 0.016 R$_{\odot}$, 0.490 $\pm$ 0.011 R$_{\odot}$), stellar temperature (3977 $\pm$ 59 K, 3539 $\pm$ 59 K), transit midpoint, orbital period, and transit depth (35 mmag, 36 mmag) as priors corresponding to TOI-6303b and TOI-6330b respectively for the posterior fits. We used four chains with 1500 burn in steps followed with an additional 4500 steps to confidently fit for the planetary parameters detailed in Table~\ref{tab:planetprop}. To ensure our posteriors are well-mixed and independent, we analyze each parameter for convergence using the Gelman-Rubin statistic where an $\hat{R}$ $\sim$ 1 represents a strong convergence \citep{ford_improving_2006}. 


\begin{deluxetable*}{llcc}
\tablecaption{Derived Planetary Parameters.   \label{tab:planetprop}}
\tablehead{\colhead{~~~Parameter} &
\colhead{Units} &
\colhead{TOI-6303b}&
\colhead{TOI-6330b}
}
\startdata
\sidehead{Instrument-Dependent Parameters:}
~~~RV Trend\dotfill & $dv/dt$ (\ms{} yr$^{-1}$) \dotfill  & 0.09$^{+4.85}_{-4.94}$ & 0.01$^{+4.83}_{-4.86}$\\ 
~~~HPF RV Offset \dotfill & $\gamma_\mathrm{HPF}$ (\ms{}) \dotfill & 5.40$^{+17.2}_{-15.7}$ & -207.27$^{+15.73}_{-16.64}$\\ 
~~~RV Jitter\dotfill & $\sigma_{\mathrm{HPF}}$ (\ms{})\dotfill & 24.3$^{+34.7}_{-17.4}$ & 30.2$^{+27.7}_{-18.8}$\\
~~~Transit Depth Dilution \dotfill & $D_{\mathrm{TESS~S18}}$ \dotfill & 0.76$^{+0.39}_{-0.36}$ & 0.75 $\pm$ 0.06\\
~~~ & $D_{\mathrm{TESS~S58}}$ \dotfill & 0.45 $\pm$ 0.03 & 0.83 $\pm$ 0.03\\ 
~~~Transit Jitter \dotfill & $\sigma_{\mathrm{TESS-S18}}$ \dotfill & 0.0061 $\pm$ 0.0003 & 0.0055 $\pm$ 0.0001 \\
~~~ & $\sigma_{\mathrm{TESS-S58}}$ \dotfill & 0.00458 $\pm$ 0.00003 & 0.00863 $\pm$ 0.00006 \\ 
~~~ & $\sigma_{\mathrm{APO}}$ \dotfill & \(\cdots\) & 0.00205 $\pm$ 0.00022 \\ 
~~~ & $\sigma_{\mathrm{RBO}}$ \dotfill & 0.00003$^{+0.00025}_{-0.00003}$ & 0.00003$^{+0.00032}_{-0.00003}$ \\ 
~~~ & $\sigma_{\mathrm{Keeble}}$ \dotfill & \(\cdots\) & 0.00458$^{+0.00220}_{-0.00446}$ \\
~~~Quadratic Limb Darkening \dotfill & [u, v]$_{\mathrm{TESS-S18}}$ \dotfill & [0.58$^{+0.54}_{-0.40}$, -0.002$^{+0.43}_{-0.40}$] & [0.22$^{+0.28}_{-0.16}$, 0.04$^{+0.25}_{-0.18}$] \\
~~~ & [u, v]$_{\mathrm{TESS-S58}}$ \dotfill & [0.58 $\pm$ 0.36, 0.13 $\pm$ 0.40] & [0.75$^{+0.30}_{-0.36}$, -0.04$^{+0.46}_{-0.36}$]\\
~~~ & [u, v]$_{\mathrm{APO}}$ \dotfill & \(\cdots\) & [0.16$^{+0.19}_{-0.12}$, 0.70$^{+0.16}_{-0.23}$]\\
~~~ & [u, v]$_{\mathrm{RBO}}$ \dotfill & [0.66 $\pm$ 0.39, 0.05$^{+0.44}_{-0.39}$] & [0.98$^{+0.25}_{-0.33}$, -0.30$^{+0.41}_{-0.24}$]\\
~~~ & [u, v]$_{\mathrm{Keeble}}$ \dotfill & \(\cdots\) & [0.60$^{+0.40}_{-0.38}$, -0.03$^{+0.41}_{-0.34}$]\\
\sidehead{Orbital Parameters:}
~~~Orbital Period\dotfill & $P$ (days) \dotfill & 9.4852360 $\pm$ 0.0000162 & 6.8500246 $\pm$ 0.0000033 \\
~~~Eccentricity\dotfill & $e$ \dotfill & 0.02 $\pm$ 0.02 & 0.34 $\pm$ 0.01 \\
~~~Argument of Periastron\dotfill & $\omega$ (radians) \dotfill & 2.05$^{+0.67}_{-4.50}$ & 0.65$^{+0.04}_{-0.03}$ \\
~~~Semi-amplitude Velocity\dotfill & $K$ (\ms{})\dotfill & 1000 $\pm$ 27 & 1710 $\pm$ 30 \\
\sidehead{Transit Parameters:}
~~~Transit Midpoint \dotfill & $T_C$ (BJD\textsubscript{TDB})\dotfill & 2459901.01440 $\pm$ 0.00011 & 2459901.36114 $\pm$ 0.00021 \\
~~~Scaled Radius\dotfill & $R_{p}/R_{s}$ \dotfill & 0.175 $\pm$ 0.007 & 0.206 $\pm$ 0.003 \\
~~~Scaled Semi-major Axis\dotfill & $a/R_{s}$ \dotfill & 26.95 $\pm$ 0.79 & 25.55$^{+0.63}_{-0.59}$ \\
~~~Orbital Inclination\dotfill & $i$ (degrees)\dotfill & 89.03$^{+0.28}_{-0.20}$ & 88.41 $\pm$ 0.16\\
~~~Impact Parameter\dotfill & $b$\dotfill & 0.45$^{+0.08}_{-0.13}$ & 0.52 $\pm$ 0.04 \\
~~~Transit Duration\dotfill & $T_{14}$ (days)\dotfill & 0.1216$^{+0.0033}_{-0.0030}$ & 0.0929$^{+0.0017}_{-0.0015}$ \\
\sidehead{Planetary Parameters:}
~~~Mass\dotfill & $M_{p}$ (M$_\oplus$; M$_J$)\dotfill &  2493 $\pm$ 99; 7.84 $\pm$ 0.31 & 3179 $\pm$ 100; 10.00 $\pm$ 0.32\\
~~~Mass Ratio\dotfill & $q$ (\%)\dotfill &  1.16 $\pm$ 0.06 & 1.79 $\pm$ 0.09\\
~~~Radius\dotfill & $R_{p}$  (R$_\oplus$; R$_J$) \dotfill& 11.59$^{+0.61}_{-0.64}$; 1.034$^{+0.054}_{-0.057}$ & 10.89$^{+0.37}_{-0.35}$; 0.972$^{+0.033}_{-0.031}$ \\
~~~Density\dotfill & $\rho_{p}$ (\gcmcubed{})\dotfill & 8.82$^{+1.77}_{-1.32}$ & 13.57$^{+1.43}_{-1.33}$ \\
~~~Semi-major Axis\dotfill & $a$ (AU) \dotfill & 0.07610 $\pm$ 0.00100 & 0.05763$^{+0.00073}_{-0.00077}$ \\  
~~~Planetary Insolation & $S$ (S$_\oplus$)\dotfill &  14.28 $\pm$ 1.51 & 9.97 $\pm$ 1.11 \\
~~~Equilibrium Temperature (albedo = 0) \dotfill & $T_{\mathrm{eq}}$ (K)\dotfill & 541 $\pm$ 14 & 495 $\pm$ 14\\
~~~Equilibrium Temperature (albedo = 0.5) \dotfill & $T_{\mathrm{eq}}$ (K)\dotfill & 455 $\pm$ 12 & 416 $\pm$ 11\\
\enddata



\normalsize
\end{deluxetable*}

\begin{figure*}[!t]
\centering
\fig{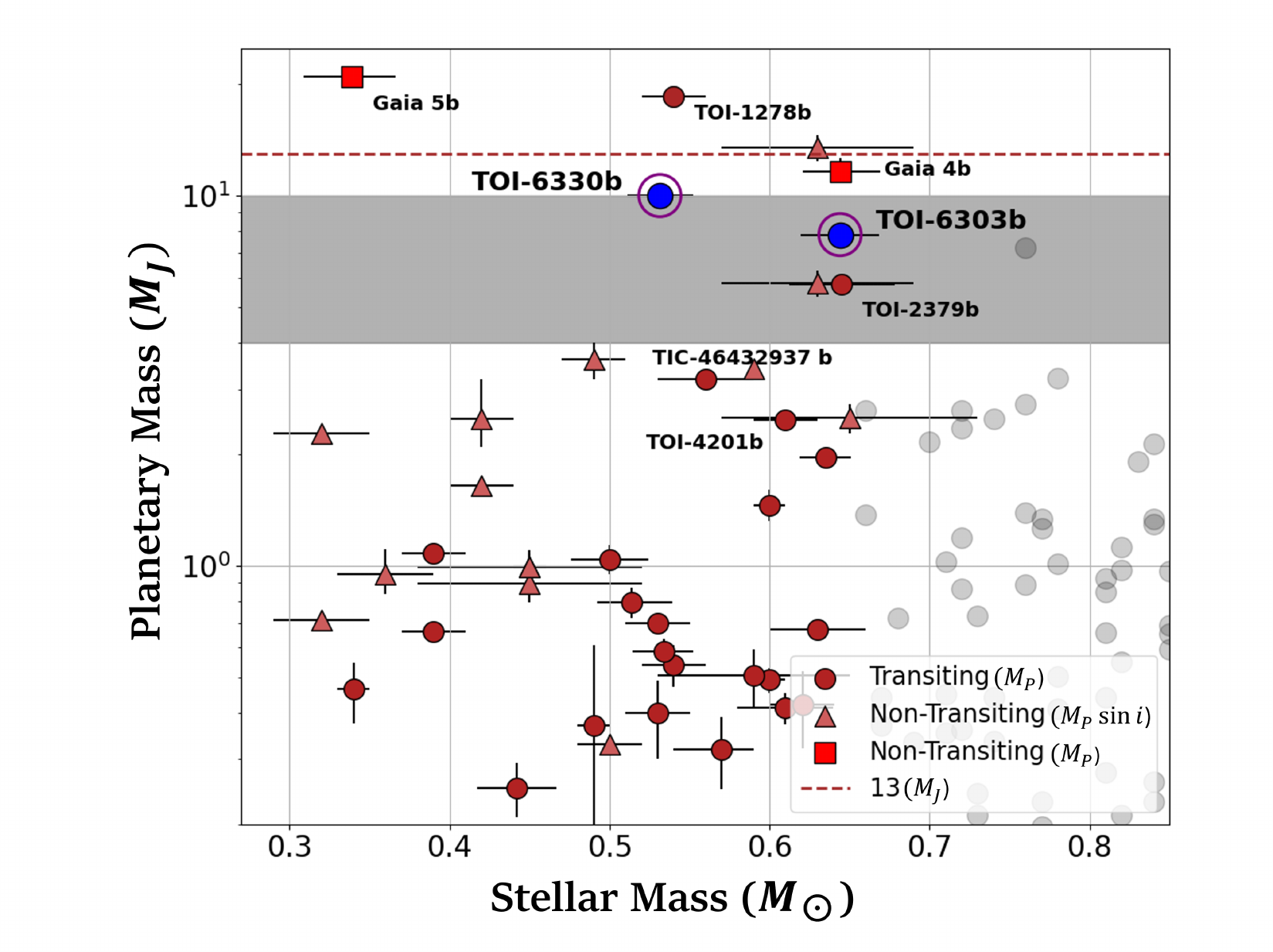}{1.0\columnwidth} {\small }
\fig{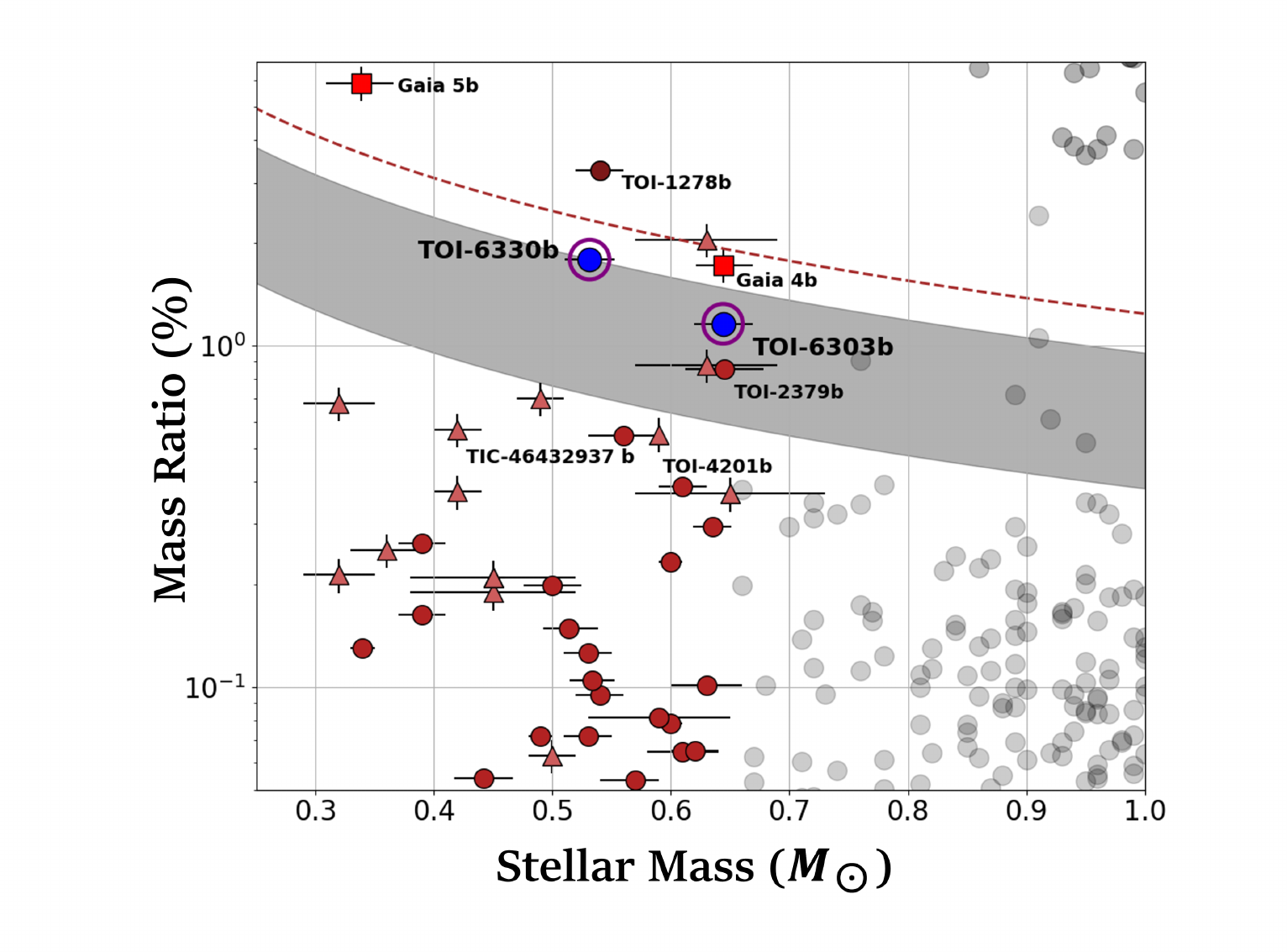}{1.05\columnwidth}
{\small }
    
\caption{\textbf{Left:} TOI-6303b and TOI-6330b (seen in blue with a ring) on a Stellar Mass vs Planetary Mass plot. The red circles designate transiting GEMS, the red triangles indicate RV-only GEMS, the grey circles indicate FGK planetary companions \citep[NASA Exoplanet Archive;][]{akeson_nasa_2013}, and the red squares are true mass-measured RV objects \citep{stefansson_gaia-4b_2024}. The grey region between 4 and 10 M$_J$ indicates a sparse region of planetary discoveries made where there is a transition between two major planetary formation paths, core-accretion and gravitational instability \citep{schlaufman_evidence_2018}. The dotted line indicates the 13 M$_J$ limit where deuterium burning occurs initiates \citep{etangs_iau_2022}. \textbf{Right:} The stellar mass vs mass ratio plot for TOI-6303b and TOI-6330b (seen in blue with a purple ring). }
\label{fig:MR_St}
\end{figure*}

 \begin{figure}[!b]
     \centering
     \includegraphics[width=0.45\textwidth]{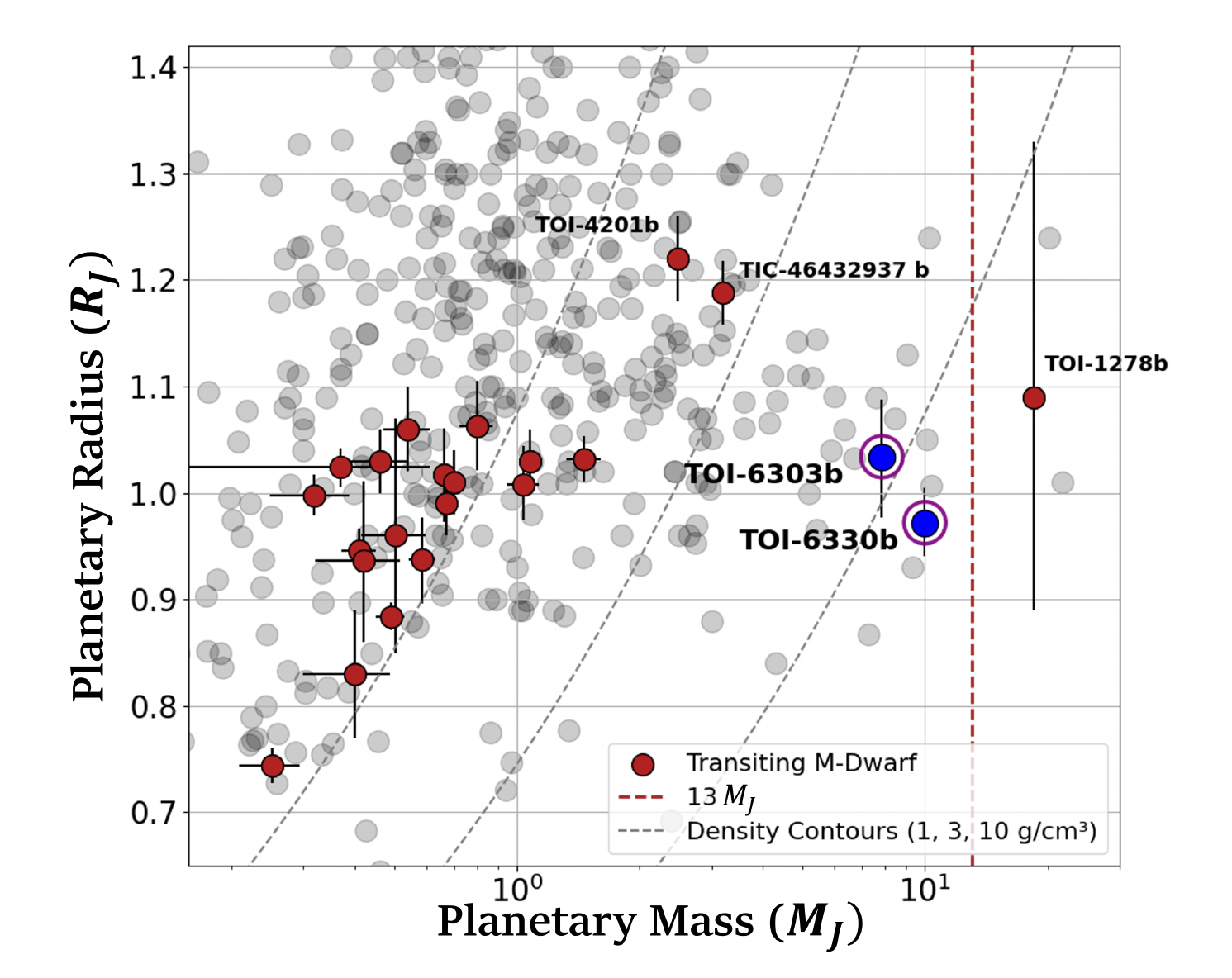}
     \caption{We show TOI-6303b and TOI-6330b (seen in blue with a purple ring) on in the planet mass-radius plane. The red circles designate transiting GEMS and the grey circles indicate FGK planetary companions. We removed NGTS-1b from this plot due to its large uncertainties on the planetary radius from its grazing transit.}
     \label{fig:Mpl_Rpl}
 \end{figure}

\section{Discussion}\label{sec:Discussion}






TOI-6303b and TOI-6330b are the most massive transiting Giant Exoplanets around M-dwarf Stars (GEMS) discovered, with masses of 7.84 $\pm$ 0.31 M$_J$ and 10.00 $\pm$ 0.32 M$_J$ mass ratios of 1.16\% and 1.79\% respectively (\autoref{fig:MR_St}). In Figure~\ref{fig:Mpl_Rpl} we plot these planets alongside other known transiting giant exoplanets and brown dwarfs around M-dwarfs. We queried NASA's Exoplanet Archive \citep{akeson_nasa_2013} on 2024 August 22 for transiting giant planets around stars with a T$_{\rm eff}$ $<$ 4100 K. We also added new discoveries K2-419Ab and TOI-6034b \citep{kanodia_searching_2024}, TOI-6383Ab (Bernabo et al. submitted), and TOI-5688Ab \citep{reji_searching_2024} from the \textit{Searching for GEMS} survey \citep{kanodia_searching_2024}, as well as the newly discovered planets of TOI-762Ab and TIC-46432937b \citep{hartman_toi_2024}. TOI-2379b is the closest planetary mass companion comparable to TOI-6303b and TOI-6330b at 5.76 $M_J$ \citep{bryant_toi-2379_2024}, while TOI-1278b is asserted to be a brown dwarf of similar mass of 18.5 $M_J$ \citep{artigau_toi-1278_2021}. 

In \autoref{fig:MR_St}, we also include recent astrometric + RV discoveries of Gaia-4b and Gaia-5b \cite{stefansson_gaia-4b_2024}, alongside radial velocity discoveries such as GJ 676 Ab \citep{forveille_harps_2011} and GJ 676 Ac \citep{sahlmann_mass_2016}. This allows us to paint a fuller picture of known giant planets around M-dwarfs. The limitation of the solely RV-measured objects discovered is that they only have a lower mass limit ($M_p$ sin \textit{i}) compared to the transit discoveries that have bonafide mass measurements ($M_p$). We compare our discoveries with super-Jupiters around M-dwarfs from different detection techniques \autoref{tab:superJupiters}.



The heavy element content of these objects has shown to be useful in offering insights into the formation of these transiting GEMS \citep[see references in][]{kanodia_toi-5205_2023, delamer_toi-4201_2024}. Using the empirical mass-metallicity relations from \cite{thorngren_mass-metallicity_2016}, we find heavy-element masses of $\sim$ 200 M$_\oplus$ and $\sim$ 235 M$_\oplus$ for TOI-6303b and TOI-6330b respectively. Note that their sample contains notable scatter and uncertainties (both systematic and statistical), such as the equations of state of H/He, sources of heating, and vertical mixing. Such a high metal-content for super-Jupiters is speculated to be a by-product of planet-planet collisions \citep{ginzburg_heavy-metal_2020}, though pinning down the exact formation and evolutionary mechanisms for TOI-6303b and TOI-6330b remains a challenge.


In the subsequent sections, we speculate on potential formation scenarios for these massive planets.

\begin{table*}[]\centering
\caption{Massive objects around M-dwarf stars in the transition between 4 M$_J$ $<$ M$_p$ $<$ 25 M$_J$ }
\begin{tabular}{ccccccc}
\hline
\hline
Object Name & M$_p$ (M$_J$) & M$_*$ (M$_\odot$) & M$_p$/M$_*$ (\%) & Period (days) & $M_p$ or $M_p$ sin\textit{i} & Reference \\
\hline
TOI-2379b & 5.76 $\pm$ 0.20 & 0.645 $\pm$ 0.033 & 0.85 & 5.469 $\pm$ 0.0000023 & $M_p$ & \cite{bryant_toi-2379_2024}\\
GJ 676 Ab & 5.792$^{+0.469}_{-0.477}$ & 0.626 $\pm$ 0.063 & 0.88 & 1051.44$^{+0.377}_{-0.400}$ & $M_p$ sin\textit{i} & \cite{forveille_harps_2011} \\
TOI-6303b & 7.84 $\pm$ 0.310 & 0.644 $\pm$ 0.024 & 1.16 & 9.485 $\pm$ 0.0000162 & $M_p$ & This work \\
Gaia 4b & 11.63$^{+0.97}_{-0.82}$ & 0.644$^{+0.025}_{-0.023}$ & 1.72 & 571.4$^{+1.5}_{-1.4}$ & $M_p$ & \cite{stefansson_gaia-4b_2024} \\
TOI-6330b & 10.00 $\pm$ 0.315 & 0.531 $\pm$ 0.021 & 1.79 & 6.85 $\pm$ 0.0000033 & $M_p$ & This work \\
GJ 676 Ac & 13.492$^{+1.046}_{-1.127}$ & 0.626 $\pm$ 0.063 & 2.04 & 13921.42$^{+1232.34}_{-1514.65}$ & $M_p$ sin\textit{i} & \cite{sahlmann_mass_2016} \\
TOI-1278b & 18.5 $\pm$ 0.5 & 0.54 $\pm$ 0.02 & 3.27 & 14.48 $\pm$ 0.00021 & $M_p$ & \cite{artigau_toi-1278_2021}\\
Gaia 5b & 20.93$^{+0.54}_{-0.52}$ & 0.339$^{+0.027}_{-0.03}$ & 5.89 & 358.57 $\pm$ 0.2 & $M_p$ & \cite{stefansson_gaia-4b_2024} \\
\hline
\end{tabular}
\tablenotetext{}{\textit{Note: We omitted discoveries via direct imaging and required all objects to have parameter values within 3$\sigma$}}
\label{tab:superJupiters}
\end{table*}



\subsection{Formation around M-dwarfs: Core-accretion}

Core-accretion is the favored formation path for close-in transiting giant exoplanets due to the observed host-star metallicity correlation \citep{fischer_planet-metallicity_2005, narang_properties_2018, osborn_investigating_2020}. It requires an initial formation of a solid core with a mass of $\sim$ 10 M$_\oplus$ to initiate a runaway gaseous accretion \citep{mizuno_formation_1980, pollack_formation_1996}. Around M-dwarf stars, the formation of gas giants through the core-accretion paradigm contains two primary issues: a prolonged formation timescale that would take too long to initiate exponential runaway gaseous accretion with respect to the gas (primarily H/He) lifetime in the disk \citep{laughlin_core_2004} and a low total dust mass in the protoplanetary disk \citep[due to the disk-to-star mass scaling][]{andrews_mass_2013, pascucci_steeper_2016}.

Super-Jupiters (2 M$_{J}$ $\le$ M$_{p}$ $\le$ 10 M$_{J}$) necessitate formation in the outer regions of the protoplanetary disk \citep[$>$ 5 AU;][]{mordasini_extrasolar_2009}. Since these objects formed around M-dwarfs, their orbital timescales are likely much longer, making planetesimal accretion much slower ($\sim$ 1 Myr), but pebble accretion remains sufficient \citep{lambrechts_rapid_2012, savvidou_how_2023}.
It is possible that young disks may coagulate more mass in their mid-plane through self-gravitating spiral waves, allowing for the faster formation of a core massive enough to initiate runaway accretion \citep{haghighipour_gas_2003, baehr_filling_2023}. Lastly, studies have shown that low-mass disks around M-dwarfs have a longer lifetime than those around more massive stars \citep{pfalzner_most_2022}.  The typical lifetime for protoplanetary disks around M-dwarfs is $\sim$ 2-3 Myr with an upper limit of $\sim$ 10-20 Myr \citep{ribas_disk_2014}. The longer living disks would provide additional time for solid cores and gaseous envelopes to form massive enough to initiate runaway accretion under the core-accretion paradigm. 

The primary issue with the core-accretion paradigm when addressing the formation of these objects is the mass limitations. Class II protoplanetary disk masses from ALMA have a large scatter in the protoplanetary disk masses \citep{pascucci_steeper_2016, ansdell_alma_2016, manara_demographics_2023}, and also suffer from systematic underestimation of disk masses \citep{miotello_setting_2022} due to complexities. Primarily, the optical thickness of protoplanetary disks cause a misrepresentation of disk masses from millimeter flux observations \citep{rilinger_determining_2023} and the hidden nature of disk structures due to an increase of opacity \citep{liu_underestimation_2022}. Furthermore, the Class II disk masses are likely not the primordial mass reservoirs for giant planets \citep{greaves_have_2010, mulders_mass_2021}, especially super-Jupiters like TOI-6303b and TOI-6330b, with giant planet formation likely beginning much earlier. 

Given the low occurrence of GEMS \citep{gan_occurrence_2023, bryant_occurrence_2023} and high mass ratios, they are clearly outliers of planet formation. Therefore to explore the feasibility of these super-Jupiters forming through core accretion, instead of considering the median disk dust masses, we estimate the maximum masses that disks can attain before getting gravitationally unstable based on simulations from Figure 7 in \cite{boss_forming_2023} as $\sim$ 10\% of the host star mass. A disk mass of 10\% the host star mass would have a total mass of $\sim$ 67 $M_J$ and 56 $M_J$ for TOI-6303 and TOI-6330 respectively. These disks are about 8.5$\times$ and 5.5$\times$ their respective planet masses, suggesting an overall (mainly gaseous accretion) formation efficiency of 12 -- 18 \%, which is an extremely unlikely efficiency rate. While a detailed hydrodynamic simulation is beyond the scope of this paper, studies have shown that the presence of such massive super-Jupiters should open up gaps in protoplanetary disks, which would reduce the gas accretion efficiency and gas mass available for accretion onto these planets \citep{1993prpl.conf..749L, 1999ApJ...514..344B, ginzburg_heavy-metal_2020}.


We can also estimate the more conventional disk dust mass budget for the massive disks assumed above, with a 1\% dust-to-gas mass ratio, to be 200 and 175 \earthmass{} for TOI-6303b and TOI-6330b respectively. In other words, the most massive disks that can stave off gravitational collapse are 10\% the host star mass, and even in this scenario, the disk dust masses are only about 100\% and 75\% the estimated heavy-element content of these planets. Conversely, estimates for the efficiency of pebble accretion efficiency are $\sim$ 10\% \citep{lin_balanced_2018}, as opposed to the 100\% and 133\% respectively, required here.

Even assuming much more favorable formation efficiency estimates would necessitate very massive disks that are $\gg$10\% the stellar mass, or uncharacteristically metal-rich. This might indeed be the case for TOI-6303, which is likely metal-rich, as opposed to TOI-6330, which is closer to Solar metallicity.

In summary, such massive planets (1.16 -- 1.79\% mass-ratios) are difficult to explain under the core-accretion paradigm without very massive disks ($>$ 10\%) that should be susceptible to gravitational collapse. This simple argument is agnostic of the limitations of protoplanetary disk mass measurements, as well as the assumptions for the epoch at which planet formation should begin.

\subsection{Formation around M-dwarfs: Gravitational Instability (GI)}

With these limitations, the gravitational instability (GI) paradigm provides an alternative mechanism, taking place earlier, in class 0/I disks or the protostellar phase \citep{kuiper_origin_1951, cameron_physics_1978, boss_giant_1997}.

Compared to previous GEMS discoveries, TOI-6303b and TOI-6330b are both a much larger percentage of their host star's mass (see Figure~\ref{fig:MR_St}). There is a sparse region of planetary discoveries these objects fall into, where, for FGK stars, we see a shift in formation paths from core-accretion to gravitational instability (GI) for objects between $\sim$ 4-10 M$_J$. Studies have shown this region exists around solar-type stars with metallicities ranging from $\sim$ $\pm$0.5 [Fe/H], which includes a noticeable trend: as the metallicities increase, there is an increase in giant planet detections \citep{fischer_planet-metallicity_2005, schlaufman_evidence_2018, maldonado_connecting_2019}. 


GI can occur in a more massive Class 0/I protostellar disk enabling it to satisfy the mass budget limitations, along with formation timescales on the order of $\sim$ 10$^3$ years \citep{boss_giant_1997}. The instability model begins within the protostellar or class 0/I disks and similarly requires $\gtrsim$ 10\% of the disk mass to initiate gravitational collapse and clump formation \citep{boss_rapid_2006, boss_formation_2011}. Fragmentation of the disk leads to self-gravitating large clumps and serves as the beginning stages of a forming giant planet \citep[For a full review of GI see][]{durisen_gravitational_2007, kratter_gravitational_2016}. The formation of a planet through GI requires cooler temperatures to allow for efficient gravitational collapse through a lack of thermal pressure, causing the collapse to occur at large distances from the host star.

\subsection{Eccentricity}

\begin{figure*}
    \centering
    \includegraphics[width=1.05\columnwidth]{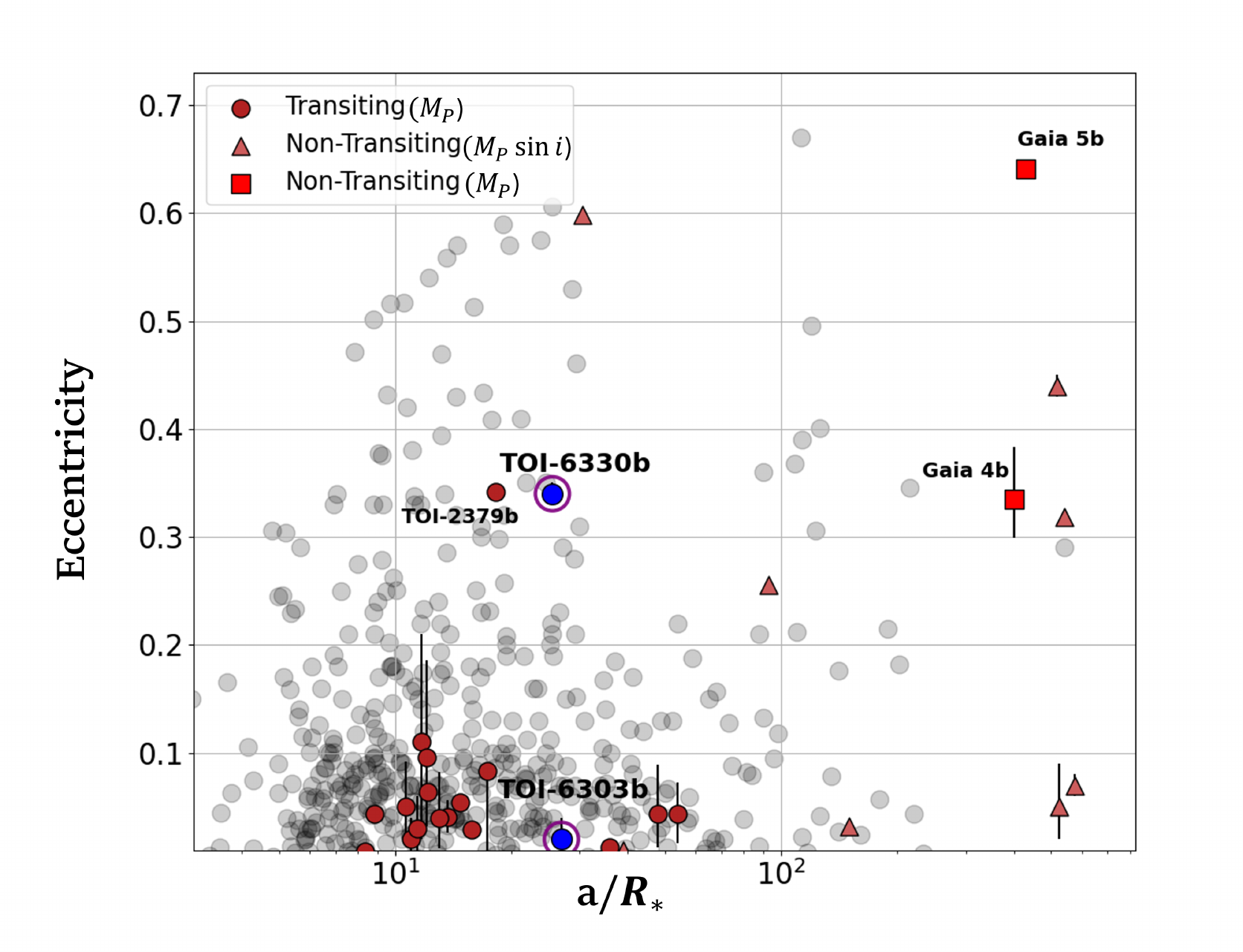}
    \includegraphics[width=1.05\columnwidth]{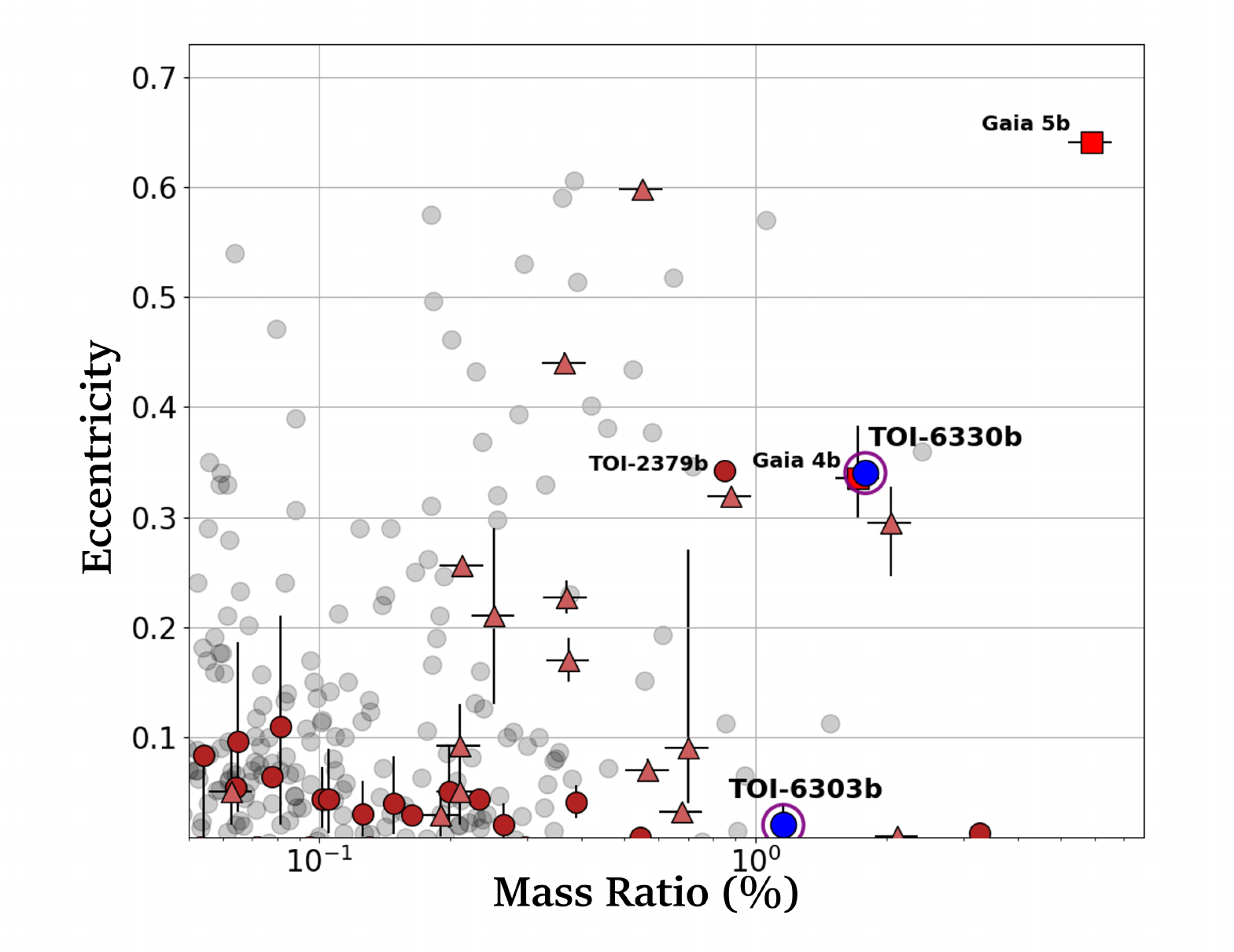}
    \caption{The semimajor axis over the Stellar Radius vs Eccentricity for TOI-6303b and TOI-6330b (seen in blue with a purple ring). The red circles indicate the transiting M-dwarf planetary companions and the grey circles are the FGK transiting companions as used in Figure~\ref{fig:MR_St}. Both TOI-6303b and TOI-6330b are at similar values for a/R$_*$ (26.95 $\pm$ 0.79 and 25.55 $^{+0.63}_{-0.59}$ respectively) but have different eccentricity values (0.02 $\pm$ 0.02 and 0.34  $\pm$ 0.01).}
    \label{fig:Ecc_AOR}
\end{figure*}

A majority of the transiting GEMS have low eccentricity values, ostensibly due to their short orbital periods and lower planetary masses leading to shorter circularization timescales (Figure~\ref{fig:Ecc_AOR}). TOI-6303b is aligned with the majority, but TOI-6330b is amongst the outliers with an eccentricity of 0.34 $\pm$ 0.01. 


We estimate the circularization timescales for TOI-6303b and  TOI-6330b using Equation 3 and 4 from \cite{persson_greening_2019} which derives the circularization timescale, $\tau_{e}$ using Equations 1 and 2 from \cite{jackson_tidal_2008}. 

\begin{equation}
    \frac{1}{\tau_e} = \frac{1}{\tau_{e,*}} + \frac{1}{\tau_{e,P}},
\end{equation}
\begin{equation}
    \frac{1}{\tau_{e,*}} = a_{P}^{-13/2}\frac{171}{16}\sqrt{\frac{G}{M_*}}\frac{R_*^5M_{P}}{Q_*}
\end{equation}
\begin{equation}
    \frac{1}{\tau_{e,P}} = a_{P}^{-13/2}\frac{63}{4}\sqrt{GM_*^3}\frac{R_{P}^5}{Q_{P}M_{P}}
\end{equation}

We assume $Q_{P}$ $\sim$ 10$^5$ and $Q_{*}$ $\sim$ 10$^7$ \citep[See][]{goldreich_q_1966, lainey_quantification_2016}. This yields a $\tau_{e}$ $\sim$ 27 Gyrs and 10 Gyrs for TOI-6303b and TOI-6330b respectively, which is much longer than the estimated age for these systems. Lower assumed $Q_P$ would make the circularization process more efficient and reduce these formation timescales. We estimate $\tau_e$ of $\sim 3$ Gyr and $\sim 1$ Gyr respectively, for an assumed $Q_P$ of $10^4$, however this value would be much lower than the assumed $Q_P$ value for Saturn with a mass of 0.3 $M_J$ of 6 $\times$ 10$^4$ \citep{goldreich_q_1966}, where $Q_P$ is expected to increase with planetary mass. While we speculate on these timescales, we also note that the actual efficiency of the processes are heavily dependent on the mass-distribution in these giant planet interiors, which are poorly constrained for such super-Jupiters. With circularization timescales larger than the age of both systems, we look to possible formation mechanisms to explain the eccentricities of TOI-6303b and TOI-6330b.



The low eccentricity for TOI-6303b (0.02 $\pm$ 0.02) is therefore suggestive of a more circular orbit from gas-disk migration \citep{kley_planet-disk_2012, dawson_origins_2018}. Conversely, TOI-6330b's migration is more indicative of high-eccentricity tidal migration \citep{beauge_multiple-planet_2012}. \cite{dawson_origins_2018} define high-eccentricity tidal migration as a two-step process where the planet migrates and is circularized through tidal interactions with the host star. This migration can appear in two facets: planet-planet scattering \citep[e.g.][]{rasio_dynamical_1996, chatterjee_dynamical_2008} and secular interactions. 


Planet-planet scattering can occur where planets form in a tightly packed system \citep{juric_dynamical_2008} or a stellar fly-by \citep{shara_dynamical_2016}, but it is unlikely that a planet obtains its low-period orbit through solely scattering. It requires many encounters for a planet to obtain its eccentricity. Additionally, an undetected planet may help maintain a high eccentricity \citep{adams_long-term_2006}. Secular interactions are the exchange of angular momentum through planets that are widely separated from one another \citep{dawson_origins_2018}. This migration takes thousands to millions of years to occur with two primary methods in swapping angular momentum: periodically \citep[e.g.][]{petrovich_hot_2015} and chaotically \citep[e.g.][]{hamers_secular_2017} with chaotically requiring three or more planets. For further discussion on both migration techniques see \cite{dawson_origins_2018}. It is important to note that we do not detect any signs of an additional companion to TOI-6330b.

Under the GI formation scenario it is also possible for the planet TOI-6330b to perturb spiral waves in the primordial marginally unstable disk \citep{paardekooper_planet-disk_2023}, the feedback from which could result in moderate eccentricities consistent with those seen here \citep[see Figure 6;][]{boss_formation_2024}. Conversely, due to the preponderance of GI to enable the formation of $>1$ compact objects (planets or brown dwarfs) in a system, the eccentricity seen here could also be ascribed to scattering events such as those described above, and seen in simulations \citep{boss_orbital_2023}.

\section{Conclusion} \label{sec:Conclusion}

We present the discovery of TOI-6303b and TOI-6330b, the two most massive transiting giant exoplanets around early M-dwarfs. These planetary-mass companions were first identified by TESS photometry and were characterized by ground-based photometry, RVs, and speckle imaging follow-up. 

TOI-6303b and TOI-6330b have mass ratios of 1.16\% and 1.79\% with their host stars respectively, requiring a formation efficiency of 12-18\% to have successfully formed through core-accretion in protoplanetary disks right on the cusp of gravitational instability (67 M$_J$ and 56 M$_J$ respectively, 10\% disk-to-star mass ratio). Furthermore, even in these massive disks, the dust masses are only 100\% and 75\% the estimated heavy-element content of TOI-6303b and TOI-6330b respectively. Even attributing the most favorable formation efficiency to these objects would require massive protoplanetary disks that would be on the cusp of gravitational collapse.

These objects share many similarities but have differing eccentricities that could provide insight into their possible formation and migration history. TOI-6303b has an approximately circular orbit, which is representative of gas-disk migration while TOI-6330b has a relatively eccentric orbit, which indicates an alternate migration technique such as planet-planet scattering or secular interactions. Our efforts through the \textit{Searching for GEMS} survey have not only helped discover and characterize interesting giant planet systems such as the ones presented in this paper, but also started to unearth trends which are starting to inform and refine our understanding of giant planet formation. 




\section{Acknowledgements}

Some of the observations in this paper made use of the NN-EXPLORE Exoplanet and Stellar Speckle Imager (NESSI). NESSI was funded by the NASA Exoplanet Exploration Program and the NASA Ames Research Center. NESSI was built at the Ames Research Center by Steve B. Howell, Nic Scott, Elliott P. Horch, and Emmett Quigley.

The Pennsylvania State University campuses are located on the original homelands of the Erie, Haudenosaunee (Seneca, Cayuga, Onondaga, Oneida, Mohawk, and Tuscarora), Lenape (Delaware Nation, Delaware Tribe, Stockbridge-Munsee), Shawnee (Absentee, Eastern, and Oklahoma), Susquehannock, and Wahzhazhe (Osage) Nations.  As a land grant institution, we acknowledge and honor the traditional caretakers of these lands and strive to understand and model their responsible stewardship. We also acknowledge the longer history of these lands and our place in that history.

These results are based on observations obtained with the Habitable-zone Planet Finder Spectrograph on the HET. We acknowledge support from NSF grants AST-1006676, AST-1126413, AST-1310885, AST-1310875,  ATI-2009889, ATI-2009982, AST-2108512, AST-2108801 and the NASA Astrobiology Institute (NNA09DA76A) in the pursuit of precision radial velocities in the NIR. The HPF team also acknowledges support from the Heising-Simons Foundation via grant 2017-0494.  The Hobby-Eberly Telescope is a joint project of the University of Texas at Austin, the Pennsylvania State University, Ludwig-Maximilians-Universität München, and Georg-August Universität Gottingen. The HET is named in honor of its principal benefactors, William P. Hobby and Robert E. Eberly. The HET collaboration acknowledges the support and resources from the Texas Advanced Computing Center. We thank the Resident astronomers and Telescope Operators at the HET for the skillful execution of our observations with HPF. We would like to acknowledge that the HET is built on Indigenous land. Moreover, we would like to acknowledge and pay our respects to the Carrizo \& Comecrudo, Coahuiltecan, Caddo, Tonkawa, Comanche, Lipan Apache, Alabama-Coushatta, Kickapoo, Tigua Pueblo, and all the American Indian and Indigenous Peoples and communities who have been or have become a part of these lands and territories in Texas, here on Turtle Island.

WIYN is a joint facility of the University of Wisconsin-Madison, Indiana University, NSF's NOIRLab, the Pennsylvania State University, Purdue University, University of California-Irvine, and the University of Missouri. 

The authors are honored to be permitted to conduct astronomical research on Iolkam Du'ag (Kitt Peak), a mountain with particular significance to the Tohono O'odham. Data presented herein were obtained at the WIYN Observatory from telescope time allocated to NN-EXPLORE (PI-Kanodia; 2023B-438370, 2024A-103024) through the scientific partnership of NASA, the NSF, and NOIRLab.

This work has made use of data from the European Space Agency (ESA) mission Gaia (\url{https://www.cosmos.esa.int/gaia}), processed by the Gaia Data Processing and Analysis Consortium (DPAC, \url{https://www.cosmos.esa.int/web/gaia/dpac/consortium}). Funding for the DPAC has been provided by national institutions, in particular the institutions participating in the Gaia Multilateral Agreement.

Some of the observations in this paper were obtained with the Samuel Oschin Telescope 48-inch and the 60-inch Telescope at the Palomar Observatory as part of the ZTF project. ZTF is supported by the NSF under Grant No. AST-2034437 and a collaboration including Caltech, IPAC, the Weizmann Institute for Science, the Oskar Klein Center at Stockholm University, the University of Maryland, Deutsches Elektronen-Synchrotron and Humboldt University, the TANGO Consortium of Taiwan, the University of Wisconsin at Milwaukee, Trinity College Dublin, Lawrence Livermore National Laboratories, and IN2P3, France. Operations are conducted by COO, IPAC, and UW.

Computations for this research were performed on the Pennsylvania State University’s Institute for Computational and Data Sciences Advanced CyberInfrastructure (ICDS-ACI).  This content is solely the responsibility of the authors and does not necessarily represent the views of the Institute for Computational and Data Sciences.

The Center for Exoplanets and Habitable Worlds is supported by the Pennsylvania State University, the Eberly College of Science, and the Pennsylvania Space Grant Consortium. 

Some of the data presented in this paper were obtained from MAST at STScI. Support for MAST for non-HST data is provided by the NASA Office of Space Science via grant NNX09AF08G and by other grants and contracts.

This work includes data collected by the TESS mission, which are publicly available from MAST. Funding for the TESS mission is provided by the NASA Science Mission directorate. 
This research made use of the (i) NASA Exoplanet Archive, which is operated by Caltech, under contract with NASA under the Exoplanet Exploration Program, (ii) SIMBAD database, operated at CDS, Strasbourg, France, (iii) NASA's Astrophysics Data System Bibliographic Services, and (iv) data from 2MASS, a joint project of the University of Massachusetts and IPAC at Caltech, funded by NASA and the NSF.

This research has made use of the SIMBAD database, operated at CDS, Strasbourg, France, 
and NASA's Astrophysics Data System Bibliographic Services.

This research has made use of the Exoplanet Follow-up Observation Program (ExoFOP; DOI: 10.26134/ExoFOP5) website, which is operated by the California Institute of Technology, under contract with the National Aeronautics and Space Administration under the Exoplanet Exploration Program.

CIC acknowledges support by an appointment to the NASA Postdoctoral Program at the Goddard Space Flight Center, administered by USRA through a contract with NASA.

\facilities{\gaia{}, HET (HPF), WIYN 3.5 m (NESSI), Shane (ShARCS), RBO, Keeble, APO/ARCTIC, ZTF, \tess{}, Exoplanet Archive}
\software{
\texttt{ArviZ} \citep{kumar_arviz_2019}, 
AstroImageJ \citep{collins_astroimagej_2017}, 
\texttt{astropy} \citep{robitaille_astropy_2013, astropy_collaboration_astropy_2018},
\texttt{barycorrpy} \citep{kanodia_python_2018}, 
\texttt{celerite2} \citep{foreman-mackey_fast_2017, foreman-mackey_scalable_2018},
\texttt{eleanor} \citep{feinstein_eleanor_2019},
\texttt{EXOFASTv2} \citep{eastman_exofastv2_2019},
\texttt{exoplanet} \citep{foreman-mackey_exoplanet-devexoplanet_2021, foreman-mackey_exoplanet_2021},
\texttt{HPF-SpecMatch} \citep{stefansson_sub-neptune-sized_2020},
\texttt{HxRGproc} \citep{ninan_habitable-zone_2018},
\texttt{ipython} \citep{perez_ipython_2007},
\texttt{lightkurve} \citep{lightkurve_collaboration_lightkurve_2018},
\texttt{matplotlib} \citep{hunter_matplotlib_2007},
\texttt{numpy} \citep{harris2020array},
\texttt{pandas} \citep{mckinney-proc-scipy-2010},
\texttt{photutils} \citep{larry_bradley_2020},
\texttt{pyastrotools} \citep{kanodia_2023},
\texttt{PyMC3} \citep{salvatier_probabilistic_2016},
\texttt{scipy} \citep{ virtanen_scipy_2020},
\texttt{SERVAL} \citep{zechmeister_spectrum_2018},
\texttt{Theano} \citep{the_theano_development_team_theano_2016},
}

\bibliography{ref, MyLibrary}

\end{document}